\documentclass[a4paper,10pt]{article}

\usepackage[round,authoryear,sort]{natbib}
\bibpunct{(}{)}{,}{a}{,}{,~}

\title{Boltzmann's $H$-theorem, its limitations, and the birth of (fully) statistical mechanics}

\author{ Harvey R. Brown\\
Faculty of Philosophy, University of Oxford\\ 10 Merton Street, Oxford OX1
4JJ, U.K.\\{\em harvey.brown@philosophy.ox.ac.uk}\\
and\\
Wayne Myrvold\\
Department of Philosophy, Talbot College\\
University of Western Ontario, 
London, Ontario, Canada N6A 3K7\\
{\em wmyrvold@uwo.ca}}

\begin{document}

\maketitle

\textit{I say that if that proof} [of the $H$-theorem] \textit{does not somewhere or other introduce some assumptions about averages, probability, or irreversibility, it cannot be valid.} Culverwell (1894b)

\bigskip
\textit{So there is a general tendency for H to diminish, although it may conceivably increase in particular cases. Just as in matters political, change for the better is possible, but the tendency is for all change to be from bad to worse.} Burbury (1894a)

\bigskip
\textit{The first man to use a truly statistical approach was Boltzmann [in 1877] and at that point kinetic theory changed into statistical mechanics even though it was another twenty odd years before Gibbs coined the expression.} ter Haar (1955), pp. 296-7.

\begin{abstract}
A comparison is made of the traditional Loschmidt (reversibility) and Zermelo (recurrence) objections to Boltzmann's $H$-theorem, and its simplified variant in the Ehrenfests' 1912 wind-tree model. The little-cited 1896 (pre-recurrence) objection of Zermelo (similar to an 1889 argument due to Poincar\'{e}) is also analysed. Significant differences between the objections are highlighted, and several old and modern misconceptions concerning both them and the $H$-theorem are clarified. We give particular emphasis to the radical nature of Poincar\'{e}'s and Zermelo's attack, and the importance of the shift in Boltzmann's thinking in response to the objections as a whole.
\end{abstract}\

\section{Introduction}

Recent years have seen a growth of studies in the foundations of classical statistical mechanics, and its historical origins. Prominent amongst these studies are the papers written by Jos Uffink dealing with \textit{inter alia} the work of Ludwig Boltzmann and his critics.\footnote{See Uffink (2004) and (2007).} Our aim in this paper is to supplement Uffink's magisterial treatment with some further observations on the nature of Boltzmann's $H$-theorem and its discontents. We follow Paul and Tatiana Ehrenfest in using their 1912 wind-tree model to bring out the logic and limitations of the theorem. We are particularly interested in thee aspects of this story. One is the nature and comparative strengths of the original objections. These include one of Zermelo's 1896 objections that is similar to Poincar\'{e}'s relatively little-known 1889 argument based on functional considerations which predated his recurrence theorem. The second is the hard-line nature of the attack coming from both Poincar\'{e} and Zermelo, each of whose antipathy towards the the kinetic theory of gases was based on considerations that went far beyond problems within the $H$-theorem. The third---and one also emphasized in Uffink's work--- is the moment of the introduction of probabilistic elements in the debate, at least on the part of Boltzmann, where by probability we mean an appeal to considerations that transcend assumptions about the actual, as opposed to expected, configuration or evolution of a single gas.

The scheme of the paper is as follows. We start in sections 2 and 3 with treatments of the original 1872 $H$-theorem for colliding molecules and the corresponding result in the 1912 wind-tree model. Section 4 contains a discussion of the history of the initial response to the $H$-theorem, with particular emphasis on the radical nature of the (remarkably similar) critiques of the kinetic theory of gases due to Poincar\'{e} and Zermelo. In section 5, the initial objections are analyzed in more detail, and their relative strengths assessed. This is followed in section 6 by a return to the history of our subject, and in particular to the famous debate amongst British commentators (as well as Boltzmann himself) that occurred in the mid-1890s in \textit{Nature} concerning the meaning of the $H$-theorem; some critical remarks are made concerning Huw Price's 1996 analysis of the role of S. H. Burbury in the debate. In section 6, various misunderstandings in the literature surrounding the critical assumption in the theorem, the famous  \textit{Stosszahlansatz}, are cleared up. Then, in section 7, we return to Boltzmann and his response to the reversibility objection, and highlight the change in his interpretation of the $H$-theorem that marked some, but by no means all, of his writings from 1877 on.

\section{Boltzmann's 1872 $H$-theorem}

\subsection{The theorem}

The $H$-theorem was based on a model of a gas consisting of $N$ hard, spherical molecules of a single species, in a container with perfectly elastic walls.\footnote{Boltzmann started his 1872 proof by stating that the particles were punctiform, but in practice treated them as if they were hard spheres; see Boltzmann (1872).} The gas is sufficiently dilute that only binary collisions need be taken into account in the dynamics. Let the distribution (or density) function $f(\mathbf{r}, \mathbf{v}, t)$ be defined such that $f(\mathbf{r}, \mathbf{v}, t)\mbox{d}^3r\mbox{d}^3v$ is the number of molecules within a volume element $\mbox{d}^3r$ about $\mathbf{r}$ and with velocity lying within the velocity-space element $\mbox{d}^3v$ about $\mathbf{v}$.\footnote{In fact, Boltzmann's 1872 density function was defined relative to kinetic-energy space, for which the $H$-function takes a slightly more complicated form than that given below.} Boltzmann's \textit{transport equation} determines how $f(\mathbf{r}, \mathbf{v}, t)$ evolves in time.

In deriving the transport equation, Boltzmann assumed that for the initial state of the gas, the momentum distribution is isotropic: $f_0(\mathbf{r}, \mathbf{v}, t) = f_0(\mathbf{r}, v, t)$. He also assumed that no external forces act on the gas and that at all times the distribution is independent of position (homogeneous): $f(\mathbf{r}, \mathbf{v}, t) =  f(\mathbf{v}, t)$. (These conditions severely limit the possible initial states of the gas. In particular, the last condition means that the $H$-theorem does not apply in the proverbial case where all the molecules of the gas are initially concentrated in a corner of the container.\footnote{In his 1872 paper Boltzmann does discuss a possible generalization of the $H$-theorem which allows for intitial spatial inhomogeneities, and is aware, as von Plato (1994) has noted, that in this case the theorem will not be universally valid. However, as Uffink (2004, 2007) has stressed, Boltzmann did not appear to think in 1872 that any exceptions could exist to his original theorem; see section 7 below.} For purposes that will later become clear, we note that Boltzmann stresses that once spatial homogeneity is established, it must persist forever in the absence of external disturbances.) It follows that the expression for $\partial f/\partial t$ depends only on collisions, so that the transport equation can be expressed as a \textit{balance equation}, in which losses are subtracted from gains during collisions. But the crucial assumption that Boltzmann made (as Maxwell did before him in 1867), later known as the \textit{Stosszahlansatz} (`assumption about the number of collisions', now sometimes abbreviated to SZA, a convention we follow in this paper), is this. 

Suppose we have two groups of molecules heading towards each other. Choose one of the groups and call it the \textit{target} group. Consider the collection $S$ of all the spatial volumes (cylinders) swept out by the target molecules in time $\Delta t$ such that if any molecule of the second group is found in $S$ there is bound to be a collision. Note that $S$ depends on the relative velocities of the molecules in the two groups; it is not defined solely in relation to the target molecules! Then suppose that the number of collisions that actually occur is the total volume associated with $S$ multiplied by the density of molecules of the second, \textit{colliding} group. This holds \textit{only if the spatial density of the colliding molecules in $S$ is the same as in any other part of space}.

This is the way the Ehrenfests expressed the SZA in their famous 1912 review article\footnote{Ehrenfest and Ehrenfest (1990), pp. 5, 6.}, and a special case will be seen in the next section. However, at this point some care needs to be taken to spell out the nature of the assumption. The target region $S$ is, to repeat, defined in relation to the relative velocities of the molecules from the two groups, so since the distribution function $f(\mathbf{r}, \mathbf{v}, t)$ is already assumed to be position independent, the assumption depends on the relationship between the momenta of the target and colliding molecules before collision. Indeed, it is tantamount to the claim that the density $F(v_1, v_2, t)$ of pairs of molecules which are about to collide within the period $[t, t+\Delta t]$ with velocities $v_1$ and $v_2$ is given by the product of the single molecule densities: 
\begin{equation}
F(v_1, v_2, t) = f(v_1, t)f(v_2, t),
\end{equation}
and thus the molecules entering into (but not out of!) collision are uncorrelated in velocity. In 1872, Boltzmann  did not indicate that the SZA was a hypothesis, rather than an automatic property of the state of a gas; the condition is introduced with neither justification nor name.\footnote{See in this connection Ehrenfest and Ehrenfest (1990), p. 84, footnote 49 and Uffink (2007), section 4.2.3. Within the literature on the kinetic theory of gases, the SZA is sometimes taken to hold for arbitrary pairs of molecules, not just those entering into collisions. See, for example, Jeans (1925), p. 17. In the present paper we shall restrict it to pairs of colliding molecules; the importance of doing so will be seen in section 7 below.} 

Boltzmann further introduced the $H$-functional (denoted $E$ in the 1872 paper, but later widely called $H$, including by Boltzmann):
\begin{equation}
H[f_t] = \int f(\mathbf{r}, \mathbf{v}, t)\ln f(\mathbf{r}, \mathbf{v}, t)\mbox{d}^3r\mbox{d}^3v
\end{equation}
which is large for narrow distributions and small for wide ones. Since the number of molecules in a gas is so large, Boltzmann treated the density function as continuous and differentiable. Now given the transport equation, and assuming SZA holds at time $t_0$, it can be shown that \begin{equation}
\frac{\mbox{d}H}{\mbox{d}t} \leq 0.
\end{equation}
Boltzmann assumed that SZA holds at all times, thus ensuring the monotonic behaviour of $H$ over time. Equality in (3) holds when the gas reaches equilibrium (the Maxwell-Boltzmann distribution, which is the unique stationary distribution consistent with the assumptions). This is \textit{Boltzmann's $H$-theorem}, one of whose assumptions we take to be the validity of SZA at all times.\footnote{Recent treatments of the $H$-theorem are found in Emch and Liu (2002) and Zeh (2001).  Zeh's discussion of the  SZA is slightly confusing. He says "Boltzmann proposed his \textit{Stosszahlansatz} (collision equation)" (p. 42) and interprets this to be the transport equation mentioned above. But Boltzmann did not use the term \textit{Stosszahlansatz}; it was introduced by the Ehrenfests in 1912 and for a quite distinct assumption, as we have seen. The significance of this no-correlations assumption is not highlighted either in Zeh's treatment, or in the recent, brief account of the $H$-theorem found in Albert (2000). Albert regards Boltzmann's homogeneity condition mentioned above, \textit{viz.} that the distribution function $f$ does not depend on position, as  ``a bit less innocent, a bit less obviously true'' (p. 55) than another of Boltzmann's assumptions, namely that the collision cross section for pairs of molecules is rotationally symmetric. But recall that Boltzmann was at first only interested in deriving equilibration only with respect to the velocity (actually, kinetic energy), not spatial, distribution, and in the process deriving the Maxwell-Boltzmann distribution. As for the reason that Boltzmann managed to get a time-asymmetric result, in so far as Albert gives one, it appears to be that low values of $H$ correspond to states with ``by far the largest number of microdestinations'' (footnote 17, p. 55). But such combinatorial considerations are quite foreign to the spirit of the $H$-theorem, and were only entertained by Boltzmann in 1877, as we shall see.} The theorem gives a mechanical underpinning to both the spontaneous, monotonic approach to equilibrium\footnote{The claim in classical thermodynamics that an isolated system found outside of equilibrium will spontaneously and irreversibly approach a unique equilibrium state is not to be confused with the second law; the disanalogies between the $H$-theorem and the second law were clarified in Uffink (2004, 2007).} and its associated entropy increase, $H$ playing the role of the negative of the entropy (`negentropy'). It is important to recognize that, unlike in phenomenological thermodynamics, negentropy is now defined for non-equilibrium configurations of the system, as well as for the equilibrium state.

As is well known, important limitations associated with the $H$-theorem were brought to light in the early  ``reversibility'' and ``recurrence'' objections. But Uffink has recently stressed\footnote{Uffink (2007), section 6.4.} that a challenge raised by O. E. Lanford in the 1970s is ``of no less importance'' than these famous objections, and it concerns the very consistency of the Boltzmann equation lying at the heart of the theorem. The problem is to demonstrate the (at least approximate) equivalence of two predictive processes. In the first, the microstate of the gas at some initial time is allowed to evolve by way of Hamiltonian dynamics for a specified interval, and the distribution function inferred from the final microstate. In the second, the Boltzmann equation is solved for the distribution function associated with the initial microstate, to determine the distribution function at the end of the specified interval.\footnote{The appropriateness of the regarding the Boltzmann equation as deterministic was also questioned by Kuhn (1978), p. 45.} Uffink has argued that the solution to this problem offered by Lanford and others sheds important light on the meaning of the $H$-theorem and the role of the SZA. However, in the present paper, we restrict ourselves largely to the original objections, and the attempt Boltzmann made to deal with them. And before we proceed, we need to look at one final aspect of Boltzmann's 1872 reasoning.

\subsection{The role of probability in the theorem, or lack thereof}

As we noted above, Boltzmann justified the well-behavedness of the density function $f$ by appealing to the large number of molecules in question. But starting with the 1899 work of Burbury\footnote{Burbury (1899).}, it has been recognized that because the number of molecules is strictly finite, $f$ can only be a well-defined continuous function in the context of a probabilistic reading of it. For instance, in his 1916 treatise on the kinetic theory of gases, Sir J. H. Jeans insisted that $f(\mathbf{r}, \mathbf{v}, t)\mbox{d}^3r\mbox{d}^3v$ is the \textit{expected} number of molecules \textit{selected at random} within a volume element $\mbox{d}^3r$ about $\mathbf{r}$ and with velocities lying within the momentum-space element $\mbox{d}^3v$ about $\mathbf{p}$.   Jeans wrote that ``no exception can be taken either to its intelligibility or truth'' in regard to this statement, precisely because of the introduction of the probabilistic notion of ``expectation''.\footnote{See Jeans (1954), sections 11 - 14. Further support for this line of thinking is found in Eggarter (1973) footnote 5 (see further references therein), and Zeh (2001), pp. 41-2, 45-6. A separate, dynamical argument in favour of interpreting the density function as probabilistic in nature is due to Kuhn (1978), and will be discussed in section 8.1 below.}

This stance may have much to recommend it in terms of rigour, but it is far from clear that it was Boltzmann's. Despite the fact that he used the word ``probability'' in relation to the density function $f$, it seems that such a notion was far more innocent in Boltzmann's 1872 reasoning than in Jeans's. When, some years later, Boltzmann was forced to soften his $H$-theorem as the result of serious objections to it (to be rehearsed below), he would in his defense refer to the fact that he had used the term probability in his 1872 paper. But the shift in Boltzmann's post-1872 thinking, one of the central themes of the present paper, reflects what in our view (following that of Uffink\footnote{See Uffink (2007), section 4.2.}) is the fact that he did not have in mind the notion of \textit{expectation}, but rather what he saw as justifiable approximation to \textit{idealized counting}, in his treatment of 1872.\footnote{It seems, as Uffink notes (2007), section 3.2.1, that Maxwell had previously had essentially the same interpretation of the velocity distribution function.}

We want to bring out this key conceptual point by way of the study of a simple mechanical model. 
Several such models exist which expose and clarify the the core elements of  the  1872 $H$-theorem: the 1907 dog-flea and 1912 wind-tree models of the Ehrenfests, and the 1959 Kac ring model.  We shall briefly review the wind-tree model, partly because it may be less well known to philosophers.\footnote{Recent  treatments of the dog-flea model are found in Ambegaokar and Clerk (1999) and Emch and Liu (2002); the Kac ring model is discussed in detail in Schulman (1997), chapter 2, and Bricmont (1995). The wind-tree model first appeared in 1912; see Ehrenfest and Ehrenfest (1990). The expression "wind-tree" is not found in the Ehrenfests' text, but Paul Ehrenfest used it in his lectures, and it has become common in the literature. A generalization of the model  involving rotated and rotating trees is found in van Holtent and van Sarloos (1980).} And we shall provide a new proof of the monotonic behaviour of the analogue $H$ functional, in which the density function has a well-defined \textit{non-probabilistic} meaning, and is not assumed to be differentiable with respect to time.

\section{The wind-tree model}

Imagine a set of $N$ identical point particles (the wind), called ``$P$-molecules'', in motion in a  2-dimensional plane extending indefinitely in space. These molecules are non-interacting and interpenetrable and are constrained to move at the same speed $v$ in the four perpendicular directions:
\bigskip

(1) \hspace{2 mm}$\rightarrow$\hspace{5mm};  \hspace{5 mm} (2) \hspace{2 mm} $\uparrow$ \hspace{5 mm};  \hspace{5 mm} (3) \hspace{2 mm} $\leftarrow$  \hspace{5 mm} ;  \hspace{5 mm}(4) \hspace{2 mm}$\downarrow$

\bigskip

Imagine also a set of identical, \textit{immobile} ``$Q$-molecules'' (the trees) of finite size and square shape, with a random but uniform distribution in the plane, and off of which the $P$-molecules scatter. Each side of each $Q$-molecule makes an angle of 45 degrees with two of the four directions above, so as to preserve the above constraint on the directions of motion.\footnote{The wind-tree model is a special case of the so-called Lorentz model, where non-interacting particles move through a random array of stationary scatterers.}
The average distance between the $Q$-molecules is large in comparison to the length of the side of these molecules.

Suppose now that collisions between the $P$- and $Q$-molecules are elastic in respect of the former. The $Q$-molecules are unaffected by the collisions, and therefore so is the above directional constraint on the motion of the $P$-molecules. The dynamics of the collisions for the $P$-molecules is given as follows.

Let $f_i (t)$  denote the number of $P$-molecules moving in the $i$-th direction ($i$ = (1), (2), (3), (4)) at time $t$; this velocity distribution will be time-dependent.
Let furthermore $N_{12}(\Delta t)$ denote the number of $P$-molecules whose motion goes, as a result of collision, from (1) to (2) (i.e. $\rightarrow$ to $\uparrow$) in the temporal interval between $t_0$ and $t_0 + \Delta t$. (This term should have the time $t_0$ as an index, but shortly we will be making claims that hold for all $t$.) Then such molecules at $t_0$ must be found within strips $S$ (in the form of parallelograms) of length $v\Delta t$ attached to the appropriate sides of the $Q$-molecules. 

We now make the \textit{Collision Assumption} (the analogue of Botzmann's SZA) for the time $t_0$:

\bigskip
\textit{The fraction of about-to-collide $P$-molecules lying in strips $S$ of a given type equals the ratio of the total area of strips of that type to the total free area of the plane.} 

\bigskip
Specifically, $N_{12}( \Delta t) = k\Delta t f_1(t_0)$, where $k$ is a factor that depends on ... and $k\Delta t$ denotes the area ratio just mentioned, which itself will not depend on the type of strip. Thus we get the balance equation
\begin{eqnarray}
f_1(t_0 + \Delta t) &=& f_1 - N_{12}(\Delta t) + N_{21}(\Delta t) + N_{41}(\Delta t) - N_{14}(\Delta t) \\
&=& f_1 + k\Delta t (f_2 + f_4 -2f_1),
\end{eqnarray}
where all the $f$s on the right hand sides are defined at time $t_0$. Analogous equations hold of course for the remaining $f_i(t_0 + \Delta t)$.

The Ehrenfests argued that one can infer by inspection of these equations that a monotonic decrease in the the differences between the $f_i$ must ensue. But this is not correct; it is not hard to find counterexamples.\footnote{What does follow from the equations is that $|f_1 - f_3|, |f_2 - f_4|$, and $|(f_1 + f_3) - (f_1 + f_2)|$ decay monotonically to zero, but they are compatible with, say, an intitial, temporary increase in $|f_1 - f_2|$.} It does however follow from the Collision Assumption that if we define 
\begin{equation}
H \equiv \sum_{i =1}^4 f_i \ln f_i
\end{equation} 
then $H(t_0 + \Delta t) \leq H(t_0)$. A proof is found in Appendix A below, one which does not assume that $H$ is differentiable with respect to $t$ (see the remarks at the end of the last section).

If it is now assumed that the Collision Assumption holds for all $t_0 + n\Delta t$, $n = 1, 2, \ldots$, then it follows from this result that $H$ decreases monotonically until it reaches its minimal value\footnote{As Jos Uffink pointed out to us (private communication), there is nothing in this reasoning, nor indeed in the original $H$-theorem, that strictly ensures that the gas will eventually reach its minimal value of $H$ even if the Collision assumption, or SZA, holds at all times. But along with Boltzmann, we shall assume that the equilibrium state of the gas is the state of minimal $H$.} of $N(\ln N - \ln 4)$, which corresponds to the equilibrium (``Maxwell-Boltzmann'') distribution 
\begin{equation}
\bar{f}_1 = \bar{f}_2 = \bar{f}_3 = \bar{f}_4 = N/4.
\end{equation}

\section{A little history}

\subsection{Background}

It took several years after 1872, the year of his $H$-theorem, before Boltzmann was forced to admit that the treatment could not be entirely correct in its original form. 

In 1874 James Clerk Maxwell, J. J. Thomson and P. G. T. Tait were aware that the result was hard to reconcile with the time reversal invariance of the dynamics of a gas. Josef Loschmidt, a friend and mentor to Boltzmann since the latter's student days\footnote{See the recent review of Boltzmann's life in Reiter (2007).} (and actually the first to calculate Avogadro's number) alluded to this problem in a paper published in 1876\footnote{Loschmidt (1876).}, and the problem is now standardly referred to as the \textit{Loschmidt}, or \textit{reversibility}  objection (\textit{Umkehreinwand}).  The problem was independently raised in 1894 by E. P. Culverwell, who in a note to \textit{Nature}, queried a treatment of the $H$-theorem due to H. W. Watson.\footnote{Culverwell (1894a).} This note, which ended with the famous query ``Will some one say exactly what the H theorem proves?'', led to a series of comments in \textit{Nature} in the following years from Watson, Joseph Larmor, G. H. Bryan, and most importantly Samuel H. Burbury and Boltzmann himself. Arguably the most significant aspect of this debate  was the clarification by Burbury of the role of a time-asymmetric assumption in the theorem, though, as we shall see in section 6 below, the connection between this assumption and the SZA was never properly clarified by Burbury.

Although the other participants of the \textit{Nature} debate were unaware of it,  Boltzmann had already given the first version of his reply to the reversibility objection in 1877, in which he introduced probabilistic considerations hitherto absent in the story. We will return in section 8 to this important turn of events, but note in the meantime first that the 1877 papers seem to have been widely ignored for the next two decades, and second that Boltzmann's probabilistic turn was largely beside the Loschmidt point.

In 1889 Henri Poincar\'{e} published a paper attempting to show that no monotonically increasing (entropy) function could be defined in terms of the canonical variables in a theory of the $N$-body system subject to Hamiltionian dynamics.\footnote{Poincar\'{e} (1889).} The paper precedes Poincar\'{e}'s famous publication in 1890 of the recurrence theorem for isolated finite dynamical systems\footnote{Poincar\'{e} (1890).}. The 1889 paper was designed to cast doubt on the proposal that Helmholtz had raised in 1882 for modeling  apparently irreversible processes by appeal to dynamical systems with hidden, or cyclic variables, along with the separation of slow- and relatively fast-changing properties of the system.\footnote{A useful treatment of this development, and its influence on Boltzmann, is found in Klein (1973).} A related argument is found in Poincar\'{e}'s 1889 lectures on thermodynamics, published in 1892 under the title \textit{Thermodynamique}\footnote{Poincar\'{e} (1892a).}. 

In 1893, Poincar\'{e} used his 1890 recurrence theorem in a second critique of the purported mechanical underpinning of thermodynamic behaviour, in a paper apparently written for philosophers\footnote{Poincar\'{e} (1893).}. Now the focus was not just Helmholtz's work but also the kinetic theory of gases, which according to Poincar\'{e} had adopted the ``English'' approach to explaining irreversible behaviour on the basis of probabilistic considerations.\footnote{Famously, in a 1870 letter to Strutt, Maxwell wrote: ``The second law of thermodynamics has the same degree of truth as the statement that if you throw a tumblerful of water into the sea, you cannot get the same tumblerful of water out again.'' (For further details, see Klein (1973), footnote 14.) In his 1899 lectures and in his 1893 paper, Poincar\'{e} used a similar analogy in describing the ``English'' hypothesis: `` ... if one had a hectolitre of wheat and a grain of barley, it would be easy to hide this grain in the middle of the wheat; but it would be almost impossible to find it again, so that the phenomenon appears to be in a sense irreversible.'' Poincar\'{e} (1893).}

 But it was  the more detailed 1896 argument\footnote{Zermelo (1896a).} by the young Ernst Zermelo (who would later achieve further fame through his work on the foundations of set theory)  that drew Boltzmann's attention to the recurrence theorem. Zermelo seems at the time to have been unaware of Poincar\'{e}'s 1893 paper; at any rate the objection to the $H$-theorem based on this theorem is now widely known as the \textit{Zermelo}, or \textit{recurrence} objection (\textit{Wiederkehreinwand}). Thus the standard terminology privileges Zermelo over Poincar\'{e}, and indeed at first sight it might seem that  Poincar\'{e}'s and Zermelo's aims and conclusions were significantly different. Let's delve into this matter in a bit more detail.
 
\subsection{The Poincar\'{e} and Zermelo critiques}
 
In his 1889 paper related to the Helmholtz program, Poincar\'{e} had written: 
\begin{quotation}
Do irreversible phenomena lend themselves in the same manner to a purely mechanical explanation? Can one, for example, in representing the world as made up of atoms and these atoms as undergoing attractions depending only on distances, explain why heat can never pass from a cold body to a hot body? I do not believe so, and I am going to explain why the theory of the renowned physicist [Helmholtz] does not seem to me to apply to apply to phenomena of this kind.\footnote{Poincar\'{e} (1889).}
\end{quotation}
Three years later, Poincar\'{e} was to reply to P. G. Tait's rather scathing 1892 review of his book \textit{Thermodynamique}, published in \textit{Nature}. Tait had berated Poincar\'{e} for, amongst other things, failing to address the statistical nature of the second law of thermodynamics.\footnote{Tait (1892).} Poincar\'{e} replied: 
\begin{quotation}
I have completely left aside a mechanical explanation of the principle of Clausius which M. Tait calls ``the true (\textit{i.e.} the statistical) basis of the second Law of Thermodynamics.''

I have not spoken of this explanation, which by the way seems to me hardly satisfactory, because I wanted to stay completely outside of all molecular hypotheses however ingenious they might be; and in particular I passed over the kinetic theory of gases in silence.\footnote{`J'ai laiss\'{e} compl\'{e}tement de c\^{o}t\'{e} une explication m\'{e}canique du principe de Clausius que M. Tait appelle ``the true (\textit{i.e.} the statistical) basis of the second Law of Thermodynamics.'' 

Je n'ai pas parl\'{e} de cette explication, qui me para\^{i}t d'ailleurs assez peu satisfaisante, parce que je d\'{e}sirais rester compl\`{e}tement en dehours de toutes les hypoth\`{e}ses mol\'{e}culaires quelque ing\'{e}nieuses qu'elles puissent \^{e}tre; et en particuliere j'ai pass\'{e} sous silence la th\'{e}orie cin\'{e}tique des gaz.' Poincar\'{e} (1892b). Curiously, Poincar\'{e}'s last remark is not strictly true; he did mention the kinetic theory in the final section of \textit{Thermodynamique}, and here admits that through its probabilistic, or ``statistical'' reading of the laws of thermodynamics, it is the only mechanical approach that ``has any chance of success''. Poincar\'{e} (1892a), section 333, p. 450.}
\end{quotation}

Now Poincar\'{e}'s 1893 paper, which, to repeat, discusses the implications of his 1890 recurrence theorem for the kinetic theory of gases, ominously starts by referring to the  ``mechanistic conception of the universe which has seduced so many good men''. The point of the paper is actually somewhat ambiguous. Poincar\'{e} raises the ``contradiction'' between the predicted recurrence phenomenon and the expectation, based on experience, that the universe is heading inexorably towards a state of heat death. And he wonders whether anyone has noted that the English kinetic theories, which for him represent the most serious attempt to reconcile mechanism and experience, can actually ``extricate themselves from this contradiction''. What follows in the paper appears to be Poincar\'{e}'s suggestion as to how this extrication comes about, and it brings no surprises. The kinetic theory of gases may in principle be capable of accounting for the universe's initial approach to equilibrium, but Poincar\'{e} argues that the theory must predict that the universe will, ``after millions of millions of centuries" awaken from this ``sort of slumber'' and start to move away from the state of heat death.
\begin{quotation}
According to this theory, to see heat pass from a cold body to a warm one, it will not be necessary to have the acute vision, the intelligence, and the dexterity of Maxwell's demon; it will suffice to have a little patience.

One would like to be able to stop at this point and hope that some day the telescope will show us a world in the process of waking up, where the laws of thermodynamics are reversed.
\end{quotation}

It is hard not to wonder whether there is a hint of irony here on Poincar\'{e}'s part. At any rate, in the final part of his 1893 paper, he mentions ``other contradictions'' in the kinetic theories; although he does not specify what they are, he expresses doubts even as to the very consistency of the theories, in so far as they contain ``reversibility in the premises and irreversibility in the conclusion''.\footnote{It is curious that earlier in the paper, Poincar\'{e} asserts, without justification, that ``reversibility is a necessary consequence of all mechanistic hypotheses.''} His final paragraph, however, tellingly (and somewhat confusingly) returns to experiment:
\begin{quote}
Thus the difficulties that concern us [with the kinetic theories] have not been overcome, and it is possible that they never will be. This would amount to a definite condemnation of mechanism, if the experimental laws should prove to be distinctly different from the theoretical ones.
\end{quote}

A final revelation of Poincar\'{e}s views is found in his 1898 article (an English translation of which was published in \textit{Nature} in the same year) on the stability of the solar system. Here, Poincar\'{e} stresses that the components of the solar system are not ``fictitious'' material points beloved of the mathematicians but extended, inelastic bodies wherein tidal forces (even in the case of wholly solid bodies) produce heat which is dissipated into space. In the long run, the solar system will consequently reach a final state from which it will not deviate.\footnote{In the absence of friction due to some medium permeating space, this final state will be such that ``the sun, all the planets and their satellites, would move with the same velocity around the same axis, as if they were parts of one solid invariable body.'' Poincare (1898).} Significantly, Poincare considers the production of heat, and the irreversible nature of entropy increase, to amount to a force acting on the system that is ``complementary'' to the Newtonian force of gravity. One of the main points of his 1898 paper was to convince the reader by way of an illuminating discussion of this complementary force that not even the movements of the heavenly bodies are exempt from that instability which is ``the law of all natural phenomena''. (Stability is being defined here as the property of the system not to deviate significantly over arbitrary time from its original orbit.)

What all of these papers from 1889 to 1898 show is a clear antipathy on Poincar\'{e}'s part to the attempt to apply mechanical principles to an understanding of thermodynamical processes, including the probabilistic reasoning in the kinetic theory of gases that he attributes to Maxwell. What about Zermelo?
 
His first 1896 paper might might easily be read to have a different tone from Poincar\'{e}'s 1893 paper. Zermelo refers to the arguments by Boltzmann and Lorentz that the Maxwell-Boltzmann velocity distribution for a gas is reached as a stationary final state; he argues that on the basis of Poincar\'{e}'s recurrence theorem ``there can be no single-valued continuous function \ldots of the [micro-]states that always increases for all initial states in some region [of the phase space], no matter how small the region''.\footnote{Zermelo (1896a). p. 212.} Under a generous construal, Zermelo might be seen to be questioning not so much the program of providing a mechanical underpinning of thermodynamics \textit{per se}, as the \textit{current form} of proofs therein of the irreversible behaviour associated with spontaneous equilibration.  He insists on the ``necessity of making a fundamental modification either in the Carnot-Clausius principle or the mechanical viewpoint." His critique is directed to the ideas of the pre-1877 Boltzmann, i.e. the attempt ''to prove that the well-known velocity distribution will be reached as a stationary final state, as its discoverers Maxwell and Boltzmann wished to do''.\footnote{Zermelo is of course doing an injustice to Maxwell, and the post-1877 Boltzmann. He refers in his paper to the above-mentioned debate in \textit{Nature} on the $H$-theorem, but obviously failed to read Boltzmann's contributions, as he effectively admitted in his second 1896 paper; see below.}  Poincar\'{e}, on the other hand, is looking in his 1893 paper beyond this development, or rather back to the probabilistic ideas of Maxwell (and Tait, J, Willard Gibbs and others).

But there are hints in the Zermelo paper of a harder line, and a reading of his subsequent reply to Boltzmann's response to his 1896 paper displays his (Zermelo's) real feelings. They are, after all, close to Poincar\'{e}'s. Zermelo, after admitting that in fact he ``was not familiar with Herr Boltzmann's investigations of gas theory'' in 1893 (by which he presumably means Boltzmann's post-1872 writings), states his position quite clearly:
\begin{quotation}
As for me (and I am not alone in this opinion), I believe that a single principle [the second law] summarizing an abundance if established experimental facts is more reliable than a mathematical theorem [Poincar\'{e}'s recurrence theorem], which by its nature represents only a theory which can never be directly verified; I prefer to give up the theorem rather than the principle, if the two are inconsistent.\footnote{Zermelo (1896b). Max Planck, under whom Zermelo had studied, accepted no deviations whatever to the second law of thermodynamics. (We are grateful to Jos Uffink for this information.) In mentioning that he was not alone in his position, could Zermelo have been thinking of Planck?}
\end{quotation} 
Now of course Boltzmann, after his probabilistic turn in 1877, and in particular in his 1896 reply to Zermelo\footnote{Boltzmann (1896).}, was arguing that the theorem and the principle, suitably interpreted, are not in conflict. But Zermelo correctly complained that Boltzmann's assertion that the Maxwell-Boltzmann distribution is not the truly \textit{stationary} final state of a gas ``does not follow sufficiently clearly from his earlier writings", and, more to the point, that it does not form the basis of an analogue of the second law of thermodynamics. Zermelo cogently argued that if the probability of $H$ decreasing from a given non-equilibrium state is defined by way of the Schuetz-Boltzmann fluctuating $H$ curve (see section 8 below), then the fact that there are as many points on the curve corresponding to that non-equilibrium state which lead to increasing $H$ as those leading to a decreasing $H$ is ruinous to the program. As Uffink noted in his detailed analysis of the debate between Zermelo and Boltzmann\footnote{This probing analysis goes considerably beyond the issues we are raising here; see Uffink (2007). If there is anything in Uffink's analysis that we might quibble with, however, it is his claim that ``... Boltzmann misrepresented Zermelo's argument as concluding that the mechanical
view should be given up. As we have seen, Zermelo only argued for a dilemma between
the strict validity of the kinetic theory and the strict validity of thermodynamics. Empirical matters
were not relevant to ZermeloÕs analysis." (section 4.5.6) This reading of Zermelo's position is hard to reconcile with the quote we gave above from Zermelo (1896b), in conjunction with his inability to see how Boltzmann accounted for the thermodynamic arrow of time in any satisfactory way.}, Zermelo anticipated the now widely appreciated point that in order to make sense of Boltzmann's probabilistic picture, at least as it stood in 1896, one has to introduce an extra assumption---what is often referred to today as the \textit{past hypothesis}--- to the effect that at the beginning of time the value of $H$ for the universe was high.\footnote{See, for example, Amegaokar and Clerk (1999) and Albert (2000).} Zermelo argued furthermore that  ``... as long as one cannot make comprehensible the \textit{physical origin} of the initial state, one must merely assume what one wants to prove; instead of an explanation one has a renunciation of any explanation.''\footnote{Zermelo (1896b). It is not clear this objection would have bothered Boltzmann, who in his initial reply to Zermelo wrote: "An answer to the question---how does it happen that at present the bodies surrounding us are in a very improbable state---cannot be given, any more than one can expect science to tell us why phenomena occur at all and take place according to certain laws.'' Boltzmann (1896).} He concluded his astute critique of Boltzmann's reasoning as follows:
\begin{quotation}
I have therefore not been able to convince myself that Herr Boltzmann's probability arguments, on which [quoting Boltzmann (1896)] ``the clear comprehension of the gas-theoretic theorem'' is supposed to rest, are in fact able to dispel the doubts of a mechanical explanation of irreversible processes based on Poincar\'{e}'s theorem, even if one renounces the strict irreversibility in favour of a merely empirical one. Indeed it is clear \textit{a priori} that the probability concept has nothing to do with time and therefore cannot be used to deduce any conclusions about the \textit{direction} of irreversible processes.\footnote{Zermelo might be said to have partially anticipated the arguments found in Uffink (2007), sections 7.6 and 7.7, and Bacciagaluppi (2007), to the effect that the introduction of stochastic dynamical principles into statistical mechanics does not serve to pick out a privileged arrow of time.}
\end{quotation}

Of course the story does not stop there. In 1897, Boltzmann was to reply to Zermelo, and introduce the remarkable suggestion that the above-mentioned past hypothesis can be avoided if the direction of time for living beings is defined by entropic increase, so opposite slopes of peaks in the fluctuating $H$-curve for the universe can have time pointing in opposite directions!\footnote{Boltzmann (1897).} But this development takes us too far afield. Let us look in more detail at the early objections to the $H$-theorem.

\section{The nature of the objections}

\subsection{The Loschmidt-Culverwell reversibility objection}

In the context of the wind-tree model, imagine a second model, identical to the first above, but in which at time $t_0$ all the $P$-molecules have the the same positions but opposite velocities, so that 
\begin{equation}
f_1' = f_3; \hspace{5 mm}f_2' = f_4;\hspace{5 mm}f_3' = f_1;\hspace{5 mm}f_4'=f_2.
\end{equation}
It is obvious that $H' = H$ at $t_0$.

But the collision dynamics are time-reversal invariant, which means that if we played a film of the collisions in the original model backwards, the behaviour of the $P$-molecules in the film would be consistent with the rules of the model. Thus for every sequence of states of the `gas', the inverse sequence is also possible---including one  with negentropy  $H$ at time $t_0$. 

Specifically, suppose the initial state $s_0$ of the system at $t_0$  corresponds to all the $P$-molecules moving in direction (1). When an arbitrary time $T$  has elapsed, a new state $s_T$ will have been achieved. Now consider preparing another initial state $s'_0$  at $t_0$ such that distribution of $P$-molecules corresponds to that of $s_T$ but with all the directions of motion reversed.  After time $T$ in this new arrangement, the system arrives at a state $s'_T$ in which all the $P$-molecules are in state (3) (the reverse of (1)). 
But consider finally the situation at a time slightly before that, \textit{viz.} $t_0 + T - \Delta t$, which shall be denoted by $T_-$. Then we must have that of the possible $N_{ij}(\Delta t)$, those where $j = 1, 2, 4$ must vanish, and it is easy to show that this is inconsistent with $N_{ij}(\Delta t) = N_{ik}(\Delta t) = kf_i(T_-)$ as required by the Collision Assumption.\footnote{See Rechtman \textit{et al.} (1991).}

\subsection{Zermelo's 1896 pre-recurrence objection}

We mentioned above Poincare's 1889 critique of Helmboltz's treatment of irreversibility by way of cyclic variables. Poincar\'{e} imagines a single-valued function $S$ of the generalized coordinates and momenta of a Hamiltonian system representing the entropy of the system\footnote{As Olsen (1993) points out in his footnote 2, Poincar\'{e} (like Boltzmann) goes beyond thermodynamics in assuming that such an entropy function is meaningful outside of equilibrium.}, and shows that $S$ cannot be a non-decreasing function along orbits. What was the lesson to be drawn according to Poincar\'{e}?
\begin{quote}
We must then conclude that the two principles of increase of entropy and of least action (understood in the sense of Hamilton) are irreconcilable. \ldots it seems probable that it will be necessary to search elsewhere [i.e. outside of Lagrangian dynamics] for the explanation of irreversible phenomena and to abandon in this case the familiar hypotheses of rational mechanics from which one derives the equations of Lagrange and Hamilton.
\end{quote}
Although Poincar\'{e}'s analysis is not strictly applicable to the $H$-theorem, the quotation indicates the general aversion he had to a mechanical treatment of irreversibility. 

It is also noteworthy that Poincar\'{e} does not appeal to Liouville's theorem. Now in his first 1896 paper, Zermelo independently provided an argument against the $H$-theorem that uses the Liouville theorem but otherwise is similar in spirit to Poincar\'{e}'s 1889 argument, and makes no appeal to the full recurrence result. In order to put Zermelo's argument---which we feel deserves attention\footnote{Olsen (1993) gives a systematic treatment of a related result due to Misra and Prigogine from \textit{c.} 1980, but seems to have overlooked Zermelo's argument.}---in modern language, we need to take a brief diversion through some familiar details.

\subsubsection{The phase space with measure-preserving flow}

Suppose we have a gas consisting of $N$ particles. Consider the space whose points represent possible  (micro-) states of the gas, specified jointly by the configuration variables $q_1 , q_2 , \ldots, q_N$ and momentum variables $p_1 , p_2 , \ldots, p_N$. We denote this $2N$-dimensional space by $\Gamma$, and the state of the gas at time $t$ by the vector $\left(\mathbf{q}(t), \mathbf{p}(t)\right) \in \Gamma$. 
Given the Hamiltonian equations of motion for the positions and momenta of the $N$ molecules, the Liouville theorem establishes that the \textit{volume} in the phase space $\Gamma$, as defined by the standard Lebesgue measure, is preserved. Any (measurable) subset of $\Gamma$ evolves under Hamiltonian flow like an incompressible fluid.

Now if the system is isolated, not all the phase space $\Gamma$ is available to it, since the motion is constrained by the conservation of energy. The phase point is confined to the hypersurface $\Gamma_E$ defined by $H(\mathbf{q},\mathbf{p}) =  E$, $H$ being the Hamiltonian of the gas. Taking a set $A$ which lies on the surface, its time-development is also on the surface. But in this case what is preserved is the measure
\begin{equation}
\mu(A) = \int_{A}\frac{\mbox{d}\sigma}{\|\nabla H\|}
\end{equation}
 where $\mbox{d}\sigma$ is the element of surface area on the energy surface induced by the Lebesque measure, and $\| \nabla H \|$ is the Euclidean vector norm of the gradient of $H$.\footnote{See Khinchin (1949), Chapter II, section 7.} What is important for our purposes is not the form of the measure, but \textit{that such a time-preserved measure exists}.

So let us consider a \textit{dynamical system} defined  by the quadruple $\langle \Gamma, \mathcal{A}, \mu, T_t \rangle$, where $\Gamma$ is the state space of the system, $\mathcal{A}$ a $\sigma$-algebra of measurable subsets of $\Gamma$, $\mu$ a measure on the measure space $\langle \Gamma, \mathcal{A} \rangle$, and  $T_t$ a one-parameter family of mappings $T_t: \Gamma \to \Gamma$ that preserve the measure $\mu$.

A measure-preserving flow is one which leaves the measure $\mu$ invariant in the following sense: for all $A\in \mathcal{A}$ and for all $t$
\begin{equation}
\mu \left( T_t (A) \right) = \mu \left( A \right)
\end{equation}
where $T_t (A)$ is the time-development of the set $A$, in the sense that $T_t A = \lbrace T_tx: x \in A \rbrace$. 

 It is further assumed that $T_0 = I$ (the identity) and that the semi-group property $T_sT_t = T_{s+t}$, for $s, t \geq 0$, holds. Normally it is assumed that (as in Hamiltonian dynamics) $T_t$ is  \textit{invertible}, which means  that the transformation is one-one, or bijective.

The specific case of a gas with invertible Hamiltonian flow is \emph{reversible} if for any $\left(\mathbf{q }_0, \mathbf{p}_0 \right)$ and $\left(\mathbf{q}_1, \mathbf{p}_1 \right)$ $\in \Gamma$ 
\begin{equation}
\left(\mathbf{q }_1, \mathbf{p}_1 \right) = T_{\tau}\!\left(\mathbf{q }_0, \mathbf{p}_0 \right) \iff T_{\tau}\!\left(\mathbf{ q}_1, - \mathbf{p}_1 \right) = \left(\mathbf{q }_0, -\mathbf{p}_0 \right)
\end{equation}
More generally, the reversibility condition (which is stronger than that of invertibility) is equivalent to the claim that the fundamental equations of motion are covariant (form invariant) under time reversal. This is the condition needed in the original Loschmidt-Culverwell objection to the $H$-theorem.

\subsubsection{The Zermelo measure-theoretic argument}

We are now in a position to reformulate the Zermelo argument in relation to a dynamical system  $\langle \Gamma, \mathcal{A}, \mu, T_t \rangle$, with the measure $\mu$ preserved by the flow $T_t$. Let us suppose that $S$ is indeed a non-negative entropy function on phase space, which is non-decreasing on orbits. Note that a crucial assumption is that the measure of the phase space $\Gamma$ is finite: $\mu\left(\Gamma \right) < \infty$. Consider a measurable set of states $g \in \mathcal{A}$, and its entire future development $G$ represented by
\begin{equation}
G = \bigcup_nT_{n\tau} (g),
\end{equation}
for some arbitrary finite $\tau$.
Note that $\mu(G) < \infty$. It follows if $S$ is an integrable function that the integral
\begin{equation}
\int_{T_t (G)} S \mbox{d}\mu
\end{equation}
is non-decreasing  with the increase of time. But this is inconsistent with the fact that $T_t (G)$ is contained in $G$, together with the assumption that $\mu (T_t (G)) = \mu (G)$, unless $S$ is constant along trajectories almost everywhere. 

Note that it is not assumed that $T_t$ is reversible, or even invertible. Variations of the proof are found in Appendix B below.

\subsection{The Poincar\'{e}-Zermelo recurrence objection}

Poincar\'{e}'s 1890 recurrence theorem is widely regarded as establishing that any conservative, finite mechanical system with energy bounded from above is quasi-periodic (though some care is needed in stating its implications accurately as we shall see). Thus, speaking loosely, if there is a sequence of states of the system involving decreasing negentropy, then if left to itself long enough (and for typical systems in statistical mechanics ``long'' means greater than the age of the universe) the system will ordinarily and eventually run through a series of states that are abritrarily close to the original series, with arbitrarily close values of $H$.  Apparently then, mechanical principles not only allow for behaviour that involves motion away from equilibrium, with decreasing entropy (increasing negentropy), but actually predict it in the very long run.

Note that if  the wind-tree molecules are enclosed in a box-shaped container 
of finite size, whose flat walls are either perpendicular or parallel to the the motions of any given $P$ molecule, then such a molecule that (elastically) strikes a wall of 
the container will retrace its path until it again hits a 
container wall at some other point, after which it again retraces its path to the original point on the wall. This shows that the motion 
of every individual $P$ molecule is periodic (since $P$ molecules that never strike a wall clearly have periodic motion) and 
therefore the motion of the complete system is quasi-periodic. Suppose again the initial $t_0$ state of the system corresponds to all the $P$ molecules moving in direction (1). When a time has elapsed corresponding to the 
period $\tau$ of the whole system, all 
the $P$ molecules will once more be moving in direction (1), but consider the situation at a time slightly before that, \textit{viz.} $t_0 + \tau - \Delta t$, which shall be denoted by $\tau_-$. In a variation of the argument used above in the context of the Loschmidt argument, we must have that of the possible $N_{ij}(\Delta t)$ those where $j = 2, 3, 4$ must vanish, and this is inconsistent with the requirement  that $N_{ij}(\Delta t) = N_{ik}(\Delta t) = kf_i(\tau_-)$.
So the Collision Assumption cannot hold at all times. This argument, like that of Loschmidt and Culverwell, requires that the dynamics be time reversal invariant. But appealing directly to the Poincar\'{e} recurrence theorem in this context requires weaker assumptions, weaker even than those in the Loschmidt argument and as weak as those in the Zermelo argument found in the last subsection.

\subsubsection{The recurrence theorem}

Suppose again we have a dynamical system  $\langle \Gamma, \mathcal{A}, \mu, T_t \rangle$, with the measure $\mu$ preserved by the flow $T_t$, and $\mu\left(\Gamma \right) < \infty$. 

\vspace{.2in}
\textbf{Theorem:} Almost all points in the phase space define orbits which return arbitrarily closely to the initial point. More specifically, consider an arbitrary measurable set $g$ in  the $\sigma$-algebra $\mathcal{A}$ and some arbitrary finite real number $\tau$.\footnote{Thinking of $\tau$ as being the time, in suitable units, taken for all the points in $g$ to leave $g$ under the flow makes the physical meaning of the theorem more intuitive, and makes  $\tau$ characteristic of the system. But the mathematical theorem goes through for arbitrary $\tau$.} The subset $\bar{g}$ of $g$ that contains \emph{all} the points $x$ in $g$ such that $T_t x \notin g$, for any finite $t > \tau$,   has measure zero: $\mu(\bar{g}) =0$. 

\vspace{.2in}
We are aware of two kinds of proof in the literature. The first is some variant of Zermelo's own 1896 proof, and is given (in varying degrees of rigour) in a number of modern texts. It uses the sets $G$ (the entire future development of $g$ as in section 5.2.2 above) and $G_{\tau}$, the entire future development of $T_{\tau}(g)$. It shows that $\mu(G_{\tau}) = \mu(G)$, and thus that there are states in $g$ that return to $g$ after some time $t \geq \tau$. That the set of such recurring states is dense in $g$ is then established by appeal to the continuity of the flow.

In fact, it can be shown that continuity can be replaced by the weaker condition that the flow is measurable; this is established in Appendix C below which contains a careful reworking of the Zermelo-type proof. It is worth stressing that in such a proof the semi-group property of the flow is required, but not its reversibility nor indeed even its invertibility.

The second kind of proof found in the literature\footnote{See Schulman (1997), pp. 70, 71.} starts by establishing the following lemma.

\bigskip
\textbf{Lemma.} Consider the $n$th time-development of the set $\bar{g}$ of non-recurring points, represented by $T_{n\tau} (\bar{g})$, where  $n$ is any any positive integer, and  $\tau$ is some fixed finite period of time. The sequence of such developments does not intersect itself, i.e. for every $n, m \geq 0$ and $n \neq m$, $T_{n\tau} (\bar{g}) \cap T_{m\tau} (\bar{g}) = \emptyset$.

\bigskip

The proof continues by introducing the set $\bar{G}$ formed from the union of all these developments:
\begin{equation}
\bar{G} = \bigcup_nT_{n\tau} (\bar{g}).
\end{equation}
Since all the developments  are disjoint, then the volume of the sum is the sum of the volumes:

\begin{equation}
\mu(\bar{G}) = \sum_n \mu(T_{n\tau} (\tilde{g})).
\end{equation}
But because of the measure-preserving nature of the flow, this sum must be infinite when $\mu (\bar{g}) \neq 0$, which contradicts our initial assumption that the volume of the whole available space is finite: $\mu\left(\Gamma\right) < \infty$. So $\mu (\bar{g}) = 0$.

Standardly, the lemma above is proven by appeal to the invertibility of $T_t$, a property of the flow that is not assumed in the Zermelo-type proof. In fact, the Lemma is not needed; a weaker lemma, which says that the developments  $T_{n\tau} (\bar{g})$ have intersection of zero measure, is all that is required and here invertibility is not needed. This is spelt out in Appendix D below.

It is obvious that the recurrence theorem poses a threat to the $H$-theorem: in the course of time there will be as many occasions in which $H$ increases as those in which it decreases. For a rigorous demonstration that no forever non-decreasing function $S$ on the (finite) phase space can be compatible with the recurrence result, see Zermelo (1896a) and Olsen (1993).\footnote{For a defense of the view that the recurrence theorem is not inconsistent with monotonic increase of entropy, see Mackey (1992), pp. 45, 46 and chapter 7. We note that Mackey defines entropy relative to an ensemble of systems, in contrast to the present discussion.} The recurrence theorem may be regarded as stronger than the Zermelo result described in the section 5.2.2; it demonstrates strictly irreversible behaviour is impossible even if a physically meaningful entropy-related function is not definable for non-equilibrium states.

\subsection{Comparing the reversibility and recurrence objections}

It is sometimes claimed that both objections show, in their different ways, that the $H$-theorem is incompatible with the laws of the micro-mechanics of a gas.\footnote{See for example, Sklar (1993), pp. 36, 37.} (Of course, a theorem is a theorem, and it can only be its interpretation that is incompatible with kinetic theory.\footnote{It is clear that what Sklar (\textit{ibid}) meant was Boltzmann's use of the theorem; see p. 37.}) But is this so in Boltzmann's case? Do the two objections reach the same damning verdict on his use of the theorem, differing  only in the evidence they bring to bear on the case?

What the Loschmidt objection establishes, given the symmetric nature of the dynamics, is, to repeat, that there are states of the gas such that its ensuing behaviour involves an increase in $H$.  Now what the $H$-theorem implies, assuming that the SZA is valid at all times, namely a certain monotonic behaviour in $H$, corresponds to what we actually observe. So the question is now: why doesn't the non-thermodynamic behaviour manifest itself? What the Loschmidt objection does is to demonstrate that Boltzmann's use of the  $H$-theorem is seriously \textit{incomplete}. First, there is no reason given as to why the SZA holds for pre-collision velocities rather than post-collision ones.\footnote{Price (1996), p. 40, stresses this point.} But secondly, and more to the point, 
so far there is no categorical reason to think that it could not be a contingent fact (unexplained for sure) that the SZA in its standard form holds at all times. (Indeed, this was Burbury's position in 1894-5, as we shall soon see.) In this light, the $H$-theorem, or rather Boltzmann's use of it, is not so much incompatible with the laws of micro-mechanics as \textit{ad hoc}, or at least incomplete.

But the recurrence theorem categorically rules out the possibility that the SZA can hold for all future time, at least for some initial states of the gas, does it not? It might seem, then, that  Zermelo's objection is stronger than Loschmidt's on two grounds. First, it requires the semi-group property in the dynamics, but not reversibility (which the Loschmidt argument presupposes), nor even invertibility, as we have seen. Second, the former excludes a possibility left open by the reversibility objection.\footnote{This point of view seems to be defended in Zermelo (1896a); see also Price (1996), p. 33.} But some care needs to be exercised in justifying this second claim.

Commentators sometimes either take the Poincar\'{e} theorem to establish that finite mechanical systems are bound to exhibit recurrent behaviour given enough time\footnote{\textit{ibid.}}, or at least do so with probability one.\footnote{See for example Albert (2000), p. 76.} But so far in the present analysis, no notion of probability has been introduced. In particular, no connection between the measure $\mu$ on the phase space and probability has explicitly been assumed. Is it needed? Is it not enough to establish, as the recurrence theorem does, that at least one point in any neighbourhood $A$ in the phase space corresponds to an orbit that returns to $A$ in a finite time? The answer is no, at least if the Zermelo objection is taken to be stronger than Loschmidt's. 

The existence of such recurring orbits will be of little import, if they are not shown to be \textit{typical} orbits for the kinds of systems we encounter in the world.\footnote{This remark may seem perverse, given the stupendously large Poincar\'{e} recurrence times for typical thermodynamical systems compared even with the age of the universe. Indeed this consideration has led some commentators to assert that the recurrence theorem is irrelevant to the question of the validity of the $H$-theorem. (See, for example, Zeh (2001), pp. 39, 40.) But we are adopting a  literal reading of the $H$-theorem, in which it is claimed that once the SZA holds, it holds for \textit{all future time}. Even if a more pragmatic reading of the $H$-theorem need not concern itself with the recurrence theorem, it nonetheless leaves obscure the issue as to how far into the future the validity of the theorem extends.} But for this to be the case, we need to endow the measure with a probabilistic significance, so as to conclude that the non-recurring orbits have negligible, ideally zero, probability of obtaining for real gases no matter how much time is available. We have moved away from purely mechanical considerations, and this state of affairs surely 
clouds the issue as to whether the Zermelo's objection is stronger than Loschmidt's, at least in the second sense above.\footnote{A word is in order concerning the assumption that $\mu\left(\Gamma \right) < \infty$, i.e. that the measure of the available part of the phase space is finite. Can this assumption be justified without introducing probability into the story? In the special case of a gas with Hamiltonian flow, the system is normally assumed to be contained in a finite spatial volume, and its energy is assumed to be bounded from above and from below. This is enough to ensure that the Lebesgue measure of the energy hypersurface in the phase space is finite; recall that it is the Lebesque measure that is preserved under the flow. So it seems natural to impose a generalized version of this condition for arbitrary dynamical systems before any probabilistic interpretation of the measure function $\mu$ is brought in.}

This brings us to another difference between the analyses given by Poincar\'{e} and Zermelo. In his 1893 paper, Poincar\'{e} never mentions the measure-zero issue. Here is how he describes, or rather over-simplifies, his recurrence result:
\begin{quotation}
A theorem, easy to prove, tells us that a bounded world, governed only by the laws of mechanics, will always pass through a state very close to its initial state.\footnote{Poincar\'{e} (1893).}
\end{quotation}
This is quite different from the treatment in Poincar\'{e}'s 1890 monograph, where he takes pains to give a ``precise definition'' of probability based effectively on phase-space measure, with respect to which the non-recurring cases are the exception and not the rule.

Zermelo, on the other hand, devotes space in his 1896 paper to a discussion of the ``singular'' phases which avoid sitting on (quasi-) recurrent orbits. He admits that the recurrence theorem can be considered consistent with the second law (and hence presumably the $H$-theorem) if 
\begin{quotation}
only those [singular] initial states that lead to irreversible processes are actually realized in nature, despite their smaller number, while the other states, which from a mathematical viewpoint are more probable, actually \textit{do not occur}.\footnote{Zermelo (1896a), p. 215.}
\end{quotation}
Needless to say, Zermelo rejects this possibility, and does so on a number of grounds. His principal reasons, which are far from conclusive, are two-fold. First, such selection on the part of Nature of special initial states is claimed to be contrary to the spirit of the mechanical world-view, which supposedly gives all elements of the phase space equal status \textit{qua} potential states of the system. Second, arbitrarily small perturbations of singular states lead to the "probable" states associated with recurrent orbits. Since, says Zermelo more controversially, the laws of nature do not refer to ``precise quantities or processes'', singularities of this kind ``exist only as abstract limiting cases''. It is noteworthy that Zermelo does not give the notion of probability here anything more than mathematical significance.

\section{A little more history: the \textit{Nature} debate of the 1890s}

The first responses to Culverwell's 1894 plea, based on the reversibility objection, for clarification of the status of the $H$-theorem included separate comments by Watson, Burbury and Bryan which are worth examining briefly.

Watson, whose treatment of the $H$-theorem it will be recalled had led to Culverwell's complaint, accepted the latter's point that the dynamically allowed behaviour of a gas is reversible, but flatly refused to accept that the monotonic decrease of $H$ would be altered by the reversal of the velocities of the molecules!\footnote{Watson (1894).}

The remarkable Burbury, a mathematician-barrister and Fellow of the Royal Society, for whom mathematical physics was a ``scientific recreation"\footnote{See Bryan (1913) and footnote 66 below. As Price (2003) notes, Burbury turned increasingly to mathematical physics as deafness eroded his effectiveness as a barrister; see also Bryan (1913).}, thought that Culverwell had questioned the very consistency of the $H$-theorem. He thought that Culverwell's claim was that both $H$-increasing and $H$-decreasing behaviour are simultaneously predicted on the basis of the theorem itself, an obvious contradiction. This was probably not Culverwell's point, but Burbury's analysis led to a useful (though initially confused) insight. Burbury first claimed, as he had done in other 1894 papers in the \textit{Philosophical Magazine}, that the $H$-theorem relies critically on a collisions-related assumption which he called ``condition A'', which is \textit{not} equivalent to the SZA but allowed him to provide a simplified proof of the theorem. Later in the debate, he seemed to conflate this assumption with the SZA, and Boltzmann took Burbury to mean the SZA from the outset.\footnote{Burbury (1894) wrote: ``If the collision coordinates be taken at random, then the following condition holds, viz:---For any given direction of \textit{R} [the relativity velocity of the colliding molecules] before collision, all directions after collision are equally probable. Call this condition A.'' A detailed analysis of the difference between Condition A and the SZA would take us too far afield; for more discussion see Dias (1994).} At any rate, what Burbury showed  was this. Consider the post-collision velocities of all the molecules involved in collisions occurring at a given time $t_0$, and suppose that Condition A holds in reverse for  these velocities. Then this is consistent with Condition A holding immediately before $t_0$ for the forward velocities only if the system was already in the equilibrium Maxwell-Boltzmann distribution. And so, said Burbury, ``Boltzmann's theorem can be applied to both motions [forward and reversed] only on condition that it has no effect on either.''\footnote{Burbury (1895a) p. 320} 

In the non-equilibrium case then, Burbury saw the time-asymmetrical nature of Condition A as preventing 
the $H$-theorem from being applied to both the direct and reversed motions, and hence defusing any contradiction in the theorem.\footnote{Burbury (1894).} 
 Burbury conceded that an increase in $H$ is possible at times in which Condition A fails to hold, but seems to have regarded such occurrences as rare at best. He admitted that the reason for this was unclear, but speculated that the permanent validity of Condition A may ultimately be due to interactions of the system with the environment.\footnote{See Burbury (1894), and particularly (1895b). Note that in his important 1955 review paper on the foundations of statistical mechanics, ter Haar (\textit{ibid.}) regards the fact that the SZA holds symmetrically only if the system is in equilibrium as providing a variant of the Loschmidt reversibility objection. This is an interesting stance, but it was clearly not Burbury's. The latter implicitly assumed the existence of an external arrow of time, with respect to which Condition A (hopefully) held continuously; the empirical fact that outside of equilibrium it did not hold in the reverse temporal sense was not of concern to him in 1894.} It is worth remarking that Burbury showed himself to be more comfortable with the $H$-theorem in the \textit{Nature} debate than he did privately some years later. In 1901 he wrote this in a letter to a colleague:
\begin{quotation}
Now I say that the law has never been proved at all, except as a deduction from Boltzmann's assumption A. If it has ever been proved, or attempted, will you tell me where? Tait, Watson, Maxwell, Weinstein, all make Boltzmann's assumption. Further, there is no evidence whatever for Boltzmann's assumption, except that it leads to the law of equal partition. We are in a vicious circle. The tortoise is supported on the back of the elephant, and when I ask what the elephant is supported on they say it is suspended from the tortoise.\footnote{This excerpt from the 1901 letter was published in Burbury's obituary which appeared in the Proceedings of the Royal Society in 1913, with the author (to whom the letter was originally written) referred to only as ``GHB''. In listing the Burbury obituary in our bibliography, we assume the author is G. H. Bryan, who was, like Burbury, a Fellow of the Royal Society in 1901, as well as one of the participants in the earlier \textit{Nature} debate on the $H$-theorem, as we have seen. It is also known that prior to 1901, the two men had already corresponded; see Dias (1994). (Bryan made his fame largely through his contributions to the physics of flight, but some of his earliest contributions were to thermodynamics and the kinetic theory of gases; his particular interest in the role of the partition theorem in Maxwell's work led to a joint publication with Boltzmann in 1895.) For further details of Burbury's post 1901 views on the $H$-theorem, see Price (2003).}
\end{quotation}

Before leaving Burbury's contribution to the debate, we note that in their 1911 encyclopedia article, the Ehrenfests derived a `consistency' result analogous to Burbury's but related to the Collision Assumption in the context of their wind-tree model;\footnote{See Ehrenfest and Ehrenfest (1912), footnote 65, pp.  85, 86.} in Appendix E below is found a somewhat more systematic proof of this result.\footnote{Another treatment involving a different model is found in ter Haar (1955), p. 296.})

Turning now to Bryan, his main point seems to have been that if a gas is approaching the Maxwell-Boltzmann distribution, it is practically impossible to ``project the molecules in their reversed motions with sufficient accuracy to retrace their steps for more than a very few collisions''.\footnote{Bryan 1894.} This is of course to concede (as Burbury did) that negentropy can in principle increase, but to assert it can do so in practice only for very short times, if at all---without appealing to any external influences.

In all cases, Culverwell was unimpressed.\footnote{Culverwell (1894b).} As for Watson's position, Culverwell noted that no one seemed to agree with it, and correctly concluded that indeed it would ``take away all physical meaning to the H theorem''. It is not entirely clear whether Culverwell understood the point of Burbury's attempt to rescue the consistency of the $H$-theorem. Concerning Bryan's reply, Culverwell thought his claim was question begging, and thus concluded that without the introduction of ``assumptions about averages, probability, or irreversibility'', Watson's treatment of the $H$-theorem cannot be valid.\footnote{See the quotation preceding the abstract of the present paper. We return to the issue of the probabilistic nature, or lack-thereof, of the $H$-theorem in section 7.} (The contributions of Boltzmann to the debate will be mentioned later.)

\subsection{Price's view of the debate: a critique}

We finish this section by critically examining some remarks related to the \textit{Nature} debate made by Huw Price in his justly influential 1996 book \textit{Time's Arrow and Archimedes' Point}.

Price regards Culverwell's critique of the $H$-theorem as more penetrating than Loschmidt's.
\begin{quotation}
Culverwell seems to have seen more clearly than any of his predecessors the paradoxical nature of the claim to have derived an asymmetric conclusion---\textit{even of a statistical nature}---from the symmetric foundations of classical mechanics. (our emphasis)\footnote{Price (1996), p. 31.}
\end{quotation}
Now it is difficult to reconcile this claim (in particular the reference to a possible statistical mode of reasoning) with the quotation we have just given from Culverwell (1894b). But more to the point is Price's own view of the nature of the real problem associated with the $H$-theorem. 

Price praises Boltzmann for coming to see that the problem is one of the \textit{asymmetry} involved in the theorem, and not the issue of its \textit{universal validity} (or whether it admits of exceptions). But this distinction, as will be clear from reading Price's text, only makes sense in terms of a probabilistic reading of the theorem (of which more in section 8 below). Culverwell on the other hand is addressing the original form of the $H$-theorem,  and sees the introduction of a probabilistic element as a possible solution to the problem the theorem poses. The problem itself for Culverwell, as it is for Loschmidt, is that the original proof ``would, if true, apply to a system obtained by reversing the velocities when the permanent configuration had been very nearly reached. Such a system would return its path and go further and further from the permanent configuration.''\footnote{Culverwell (1894a).}

As far as Burbury is concerned, Price first regards his comments as missing the point. Even though Burbury ``draws attention to the crucial role of the independence assumption [SZA]'', he is mistaken, according to Price, to place emphasis on the justification of the continuous validity of the SZA over time.\footnote{Price (1996), p. 31.} Now we saw above that in fact Burbury did not (at least initially in the \textit{Nature} debate) draw attention to the SZA at all, but concentrated on the role of his Condition A. However, for the sake of argument, let us (as Burbury eventually seems to have done) conflate the two conditions. The relevant question is why Price thinks Burbury's concern with the durability of the SZA over time is mistaken. The answer is that 
for Price, the real concern is with the time-asymmetric nature of the SZA whenever it holds. Note that this is not the notion of asymmetry that Burbury himself stresses, which we discussed above, and which Price does not mention. It is not, to be specific, related to the fact that the SZA cannot successively run in both temporal directions without the system being in equilibrium, in the precise sense above. Price is concerned rather with the fact that the SZA, at the time in which it is taken to hold, refers to the pre-collision state of the gas, not its post-collision condition. It looks to the future and not to the past. Burbury, though astute, is being ``sidetracked'': his worry about sustaining the SZA over time ``isn't a concern about the asymmetry reflected in [SZA] ... itself, ... for it doesn't raise any objection to the assumption that the molecules are uncorrelated before they begin to interact.''\footnote{Price (1996), p. 31.}

This is true as far as it goes, but in our view Burbury was right to stress the issue of the continuing validity of the SZA over time, in the context of the original $H$-theorem. The SZA is a constraint on the state of a gas at a given time. For an isolated gas, over times comparable to the recurrence time there will be on average at least as many instants at which it fails to hold as there are in which it holds. The recurrence theorem, to repeat, ensures that SZA cannot hold at all times. But in order to get an arrow of time of the kind Boltzmann originally wanted, there could be no exceptions, or only rare ones, to the validity of the SZA, and Burbury was right to stress this point. His speculative suggestion that it may be so because the gas is not after all strictly isolated is open to criticism, as Price himself notes.\footnote{Price (1996), p. 32.} But the suggestion also happens to be in conflict with one of the key assumptions of the recurrence theorem, namely the semi-group property of the measure-preserving flow.

It is noteworthy that in his final conclusions about the significance of the $H$-theorem, Price deliberately shifts his ground and denies that the real problem with the SZA is that it is time-asymmetric (despite, as we have seen, having berated Burbury for not being troubled by this). Now, the real problem is that SZA, and other conditions like it in the literature, \textit{fail in reverse}.
\begin{quote}
After all, if they did not fail in reverse then the [$H$-] theorem could be applied in reverse, to ``show'' that entropy increases (or $H$ decreases) toward the past, as well as toward the future. Since it is the fact that entropy does not behave in that way toward the past that actually requires explanation; it is the failure of the required assumptions in that direction that should be the focus of attention.\footnote{Price (1996), p. 40. See also Price (2006).}
\end{quote} 
There is no doubt that Price has a valid point here, but by insisting on the continual validity of SZA over time, Burbury was precisely addressing the issue Price is emphasizing. If (for whatever reason) the SZA were to hold for a significant timescale, then by Burbury's own argumentation its time reverse cannot hold in the same period unless the system is in equilibrium.

\section{When does the \textit{Stosszahlansatz} hold?}

It is striking how many distinct views have been expressed concerning the domain of validity of the SZA.

(i) It was in his third contribution to the 1890s \textit{Nature} debate that Burbury expressed his doubts most strongly about the permanent validity of the SZA, when the system is ``finite, and to be left to itself unaffected by external disturbances''.\footnote{Burbury (1895). As before, we are not distinguishing here between the SZA and Condition A, which is what Burbury actually refers to.} He reiterates his view that it is only by resorting to constant ``disturbances from without'' that the SZA will be kept ``in working order'', so as to ensure the monotonic decrease in $H$. Burbury's note elicited a reply from Boltzmann, also published in \textit{Nature}.\footnote{Boltzmann (1895b).}

Boltzmann did not agree that external disturbances are necessary to keep the SZA in working order over time. Although he accepted that such permanence was not ``a mechanical necessity'', he argued that when the system is very large and the mean (free) path of molecules is very large compared to the mean distance between two neighbouring molecules, the SZA would be perpetuated over time, at least with high probability. We return to the issue of Boltzmann's post-1872 probabilistic reading of the $H$-theorem in section 8 below; in the meantime we note that it is very hard to reconcile Boltzmann's answer to Burbury with his (Boltzmann's) notion of a time-symmetric fluctuating $H$-curve (see below).

(ii) A claim that seems to have originated in Huang's 1963 textbook on statistical mechanics, and one repeated by Davies in 1974,  is that outside of equilibrium the $H$-theorem can hold only at  times corresponding to local peaks of the $H$ curve.\footnote{See Huang (1963) and its 1987 edition, pp. 85 and 87 respectively. Huang attributes the argument to unpublished work by F.E. Low. It is repeated in Davies (1974), p. 59; one might surmise that Davies, who cites Huang's book in another context, did not discover it independently.} Huang's argument is simple: the $H$-theorem says that if at time $t_0$ the SZA holds, then $\mbox{d}H/\mbox{d}t \leq 0$ at $t_0+\epsilon$ ($\epsilon \to 0$), and it would seem to follow on grounds of symmetry that $\mbox{d}H/\mbox{d}t \geq 0$ at $t_0-\epsilon$.\footnote{Indeed in Huang (1963), p. 85, it is claimed that the $H$-theorem is actually time-reversal invariant.}

In the context of our treatment above of the $H$-theorem, Huang's conclusion would be correct if, supposing the SZA is valid at a given time, it must likewise be valid for the state of the gas obtained by reversing the velocities of all the molecules that have just been involved in collisions at that time.\footnote{Note that the issue that Huang is raising is different to that of the Ehrenfests and our Appendix F; Huang is concerned with the conditions under which SZA holds at a given time for pairs of molecules just about to collide, and its time reverse holds \textit{at the same time} for (generally different) pairs of molecules that have just collided.} This claim is far from obviously true, and Eggarter in 1973 actually constructed a model of colliding particles for which the SZA is valid at all times and its time reverse never valid.\footnote{Eggarter (1973).} However, as Eggarter noted, Huang actually assumes in his treatment of the $H$-theorem the condition of uncorrelated velocities for \textit{all} pairs of molecules of the gas, and not just those about to collide at the time in question. This ``molecular chaos'' condition is indeed time symmetric and Eggarter seems to agree with Huang that if it holds at a given time, the state of the gas will correspond to a local peak in the $H$ curve---though Eggarter stresses that now there is nothing in the $H$-theorem that establishes an average decrease of $H$ over time. By replacing the SZA with the stronger condition of ``molecular chaos'' in this sense, all hope of mechanically modelling the time-asymmetric, spontaneous approach of a gas to equilibrium is lost.

But the situation is even worse than this. What both Huang and Eggarter seemed to have overlooked is that this symmetric molecular chaos condition does not imply the SZA either in the forwards or backwards sense in time. The set of pairs of molecules about to collide (or having just collided) at the time in question may in principle exhibit correlations in their velocities even when the set of all pairs of molecules fails to do so. Thus molecular chaos does \textit{not} automatically lead to $\mbox{d}H/\mbox{d}t \leq 0$ at $t_0+\epsilon$ (nor to $\mbox{d}H/\mbox{d}t \geq 0$ for $t_0-\epsilon$).\footnote{This point was clarified in discussions with Jos Uffink.} 

(iii) It has recently been claimed by Lockwood, relying in part on the arguments given by Burbury in 1895, that the \textit{Stosszahlansatz} can only hold at equilibrium.\footnote{See Lockwood (2005), p. 208, and Burbury (1895); Lockwood wrongly states that the value of Boltzmann's $H$-function is zero at equilibrium.} But this claim is also incorrect, and based on a misreading of Burbury's logic. Lockwood argues that collisions must generate correlations between molecules; the correlations being a kind of `memory' of the collision. Time reversal invariance of the theory is then taken to imply that  pre-collision correlations must exist. Instead of concluding that the \textit{Stosszahlansatz} cannot ever hold, Lockwood concludes that it holds only when Boltzmann's $H$-function is constant, which he associates with equilibrium. There are a number of non-sequiturs in this reasoning; it is enough to note that a clear-cut refutation of Lockwood's conclusion is found in (iv) below.

(iv) In his 1973 lectures on statistical mechanics, Pauli notes\footnote{Pauli (1973), p. 8} that the \textit{Stosszahlansatz} ``picks out a specific direction of time'', and later says  of his version of the $H$-theorem:
\begin{quote}
What we have calculated by means of the Stosszahlansatz is not valid for a single gas; it is valid only for a statistical ensemble. With a statistical ensemble, the values of [$H$],\footnote{Actually, the term here is $\mathcal{H}$, which is the spatial integral of the $H$ function; Pauli's treatment does not presuppose that the distribution is position independent.} necessarily discrete for a single gas, are replaced by a continuous distribution of values which satisfies the $H$-theorem. However, when we thermodynamically observe irreversibility, the situation is always such that the distribution of positions and velocities approaches the equilibrium distribution with overwhelming probability, though small fluctuations continually occur.\footnote{\textit{ibid.}, pp. 23,4.}
\end{quote}

Pauli is suggesting that for the single gas, behaviour consistent with the $H$-theorem is actually to be understood as overwhelmingly probable, but not certain, and we return to this issue in the next section. But he seems to be suggesting that if we consider an infinite ensemble of gases, the $H$-theorem can be rescued if now defined relative to a `distribution' of values of $H$. It is not clear precisely what Pauli has in mind here. It is certainly the case that Boltzmann himself never provided a systematic statistical version of the $H$-theorem.\footnote{See Uffink (2004), section 5.3.}

A statistical simulation of the wind-tree model was provided by Rechtman \textit{et al.} in 1991.\footnote{Rechtman \textit{et al.} (1991).} These authors used two different computer programs to simulate the model, with the $P$-molecules moving discretely on a regular square lattice with periodic boundary conditions. The simulation was repeated 50 times, in each case with 2000 $P$- and 500 $Q$-molecules (the latter having sides of $2\sqrt2$) on a lattice of 45,000 sites. In each run of 1000 steps the positions of both types of molecules was re-randomized, but the initial state corresponded to all the $P$-molecules moving in direction (1) above.

In a typical run the system reached equilibrium after about 100 steps, i.e. within the first 10\% of the run. By defining suitable measures of deviation from the Collision Assumption, Rechtman \textit{et al.} were able to show that on average over the ensemble of runs, not more than 2\% of the $P$-molecules violate the condition at any step. Rechtman \textit{et al.} regarded this result as ``remarkable''.

But what does this tell us about the behaviour in a single run? Here we see (typically) a quick approach to equilibrium which must be followed, after a sufficient number (much larger than 1000) of steps, by a return to the initial state. The presupposition that the initial approach to equilibrium for a single gas gains some kind of explanation by way of the statistical analysis of the ensemble is, of course, debatable---as is the whole approach initiated by Gibbs that the behaviour of individual systems can be understood by studying the behaviour of a representative ensemble.

The study of Rechtman \textit{et al.} is consistent with the contention that the SZA can hold when the system is not in equilibrium. It may also be read as suggesting that the SZA is not a necessary condition for thermodynamic behaviour, i.e. for $H$ to decrease in the next instant. 

\section{Boltzmann's  probabilistic turn}

\subsection{Boltzmann}

Boltzmann made two contributions to the \textit{Nature} debate in the 1890s. As we saw in the previous section,  he argued \textit{pace} Burbury that the persistence of the validity of SZA over time can be justified without appeal to disturbances originating from outside the system. But the more important clarification --- and one without which Boltzmann's response to Burbury cannot be fully understood --- came in an earlier and lengthier paper, published on February 28, 1895. The arguments in this paper are now well-known. For our purposes, what is relevant is first Boltzmann's clarification that 
\begin{quote}
It can never be proved from the equations of motion alone, that the minimum function H must always decrease. It can only be deduced from the laws of probability, that if the initial state is not specially arranged for a certain purpose, but haphazard governs freely, the probability that H decreases is always greater than that it decreases. ... What I have proved in my papers is as follows: It is extremely probable that H is very near to its minimum value; if it greater, it may increase or decrease, but the probability that it decreases is always greater. Thus, if I obtain a certain value for $d$H/$dt$, this result does not hold for every time element $dt$, but it is only an average value.\footnote{Boltzmann (1895a).}
\end{quote}

The detailed probabilistic analysis of the fluctuating nature of $H$ in this paper, which Boltzmann attributes to his assistant Dr Schuetz, is essentially that given so famously and in more detail in the Ehrenfest's 1912 encyclopedia article (who reserve the term``molecular chaos'' for the probabilistic version of the SZA).\footnote{See also Price (2006), section 6.} At any rate, in his 1895 reply to Culverwell \textit{et al}., Boltzmann is reiterating the probabilistic position he adopted in his first 1877 paper\footnote{Boltzmann (1877a).} in response to Loschmidt's objection to the original form of the $H$-theorem, which had escaped the notice of his intended readers. In broad terms, Boltzmann had already done in 1877 what Culverwell was calling for in 1894 (although the extent to which the details of Boltzmann's position would have satisfied Culverwell remains unclear).

Recall that in 1872, Boltzmann gives no indication that the SZA could fail; he wrote then of his $H$-theorem
\begin{quotation}
It has thus been rigorously proved that, whatever may be the initial distribution of kinetic energy, in the course of a very long time it must always necessarily approach the one found by Maxwell.\footnote{Boltzmann (1872).}
\end{quotation} 
But from 1877 such a process of equilibration becomes for Boltzmann merely probable (and as we have seen in 1895 its persistence over time depends on the size of the gas and in particular the relative sizes of the mean free path and distances between neighbouring molecules). The change in thinking is particularly evident in the treatment of the homogeneity of the gas. For Boltzmann, in 1872, once this condition is achieved it is permanent. But in 1877, he flatly denies such permanence for arbitrary initial states. \textit{The understanding of irreversibility has taken on a new form}, despite some very misleading remarks by Boltzmann to the contrary. The significance of this shift of reasoning on Boltzmann's part cannot be overstressed; although some commentators have appreciated it, particularly the Ehrenfests in 1912, D. ter Haar in 1955\footnote{See ter Haar (1955) and the quote from this article appearing at the beginning of the present paper. As Uffink (2007) notes, ter Haar is referring to the second of the 1877 papers of Boltzmann.}, Kuhn in 1978\footnote{See Kuhn (1978), chapter II; we have comments on Kuhn's analysis below.}, and most systematically Uffink\footnote{Uffink in his (2007) is concerned not just with the real shift in Boltzmann's thinking but the confusing business of Boltzmann's own recognition of this shift in the years following 1877. Uffink warns, p. 10, that ``Boltzmann himself never indicated a clear distinction between these two different theories [kinetic theory of gases and statistical mechanics], and any attempt to draw a demarcation at an exact location in his work seems somewhat arbitrary." But Uffink \textit{ibid}, section 4, is quite categorical that in his two 1877 papers, Boltzmann's thinking underwent a major shift in relation to that in his 1872 paper.}, it seems still not to be universally understood.

For example, in their lively 2002 discussion of the $H$-theorem, Emch and Liu note, correctly, that in the light of the reversibility objection Boltzmann stressed in the mid-nineties that the $H$-theorem ``has the character of a statistical truth'', and that the second law of thermodynamics is ``from a molecular point of view, merely a statistical law.'' They write further:
\begin{quote}
The additional, somewhat hidden, assumption of statistical nature in Boltzmann's work is often referred to as Boltzmann's \textit{Stosszahlansatz}. When Boltzmann introduced it, it might have been somewhat buried in the argument, but there can be no doubt that Boltzmann later recognized the probabilistic -- i.e. non-mechanistic -- component in the theory ...\footnote{Emch and Liu (2002), p. ??. }
\end{quote}
But is there is not a confusion here? Boltzmann did not belatedly come to the conclusion in the mid-nineties that the SZA was born with an intrinsically probabilistic  pedigree. It is true that this assumption is non-mechanistic in the sense that it does not follow from the postulated dynamics of the gas molecules. And it is true that Boltzmann spoke in his original 1872 paper of the (normalized) distribution function $f(\mathbf{r}, \mathbf{v}, t)$ as a probability.\footnote{See Uffink (2004), section ??} But the SZA (and the related Collision Assumption in the wind-tree model) concerns the state of a single gas, not an ensemble of such gases, and nor does it have anything to do with our ignorance of the detailed configuration of the system when $N$ is large. To say, as Boltzmann did in 1877 and in the 1890s, that the condition is only probably the case, is not to say that the condition itself has some intrinsically probabilistic meaning. When Boltzmann, in his final 1895 response to Burbury concerning the persistence of the SZA over time, wrote that ``Condition A is simply this, that the laws of probability are applicable for finding the number of collisions'', he was using the notion of probability in a qualitatively different way to that found in his original 1872 reasoning. It was the way that Maxwell had been describing the probabilistic nature of the second law of thermodynamics already for some years, and which he would articulate clearly again in his 1878 review of P. G. Tait's \textit{Sketch of Thermodynamics}:
\begin{quotation}
If we ... consider a finite number of molecules, even if the system to which they belong contains an infinite number, the average properties of this group \ldots are still every now and then deviating very considerably from the theoretical mean of the whole system \ldots

Hence the second law of thermodynamics is continually being violated, and that to a considerable extent, in any sufficiently small group of molecules belonging to a real body. As the number of molecules in the group is increased, the deviations from the mean of the whole become smaller and less frequent; and when the number is increased till the group includes a sensible portion of the body, the probability of a measurable variation from the mean occurring in a finite number of years becomes so small that it may be regarded as practically an impossibility.

This calculation belongs of course to molecular theory and not to pure thermodynamics, but it shows that we have reason for believing the truth of the second law to be of the nature of a strong probability, which, though it falls short of certainty by less than any assignable quantity, is not an absolute certainty.

Several attempts have been made to deduce the second law from purely dynamical principles, such as Hamilton's principle, and without the introduction of an element of probability. If we are right in what has been said above, no deduction of this kind, however apparently satisfactory, can be a sufficient explanation of the second law.\footnote{Maxwell (1878), p. 280.}
\end{quotation}

Now in his elegant 1973 study of Boltzmann's debt to Helmoltz, Martin Klein regards Boltzmann's work in the early 1870s as the birth of statistical mechanics. Boltzmann had
\begin{quote}
learned from Maxwell that it was essential to describe a gas by statistical methods, that the statistical distribution of molecular velocities was basic to any analysis of the properties of a macroscopic system. Boltzmann made this approach his own, generalizing it and applying it to new problems. Foremost among these was the problem he had started with, and in 1871, Boltzmann gave a new ``analytical proof of the second law''. This time the second law, or rather those aspects of the second law that deal with equilibrium and reversible processes, appeared as theorems in a new and as yet unnamed discipline---statistical mechanics. The laws of probability played as essential a part as the laws of mechanics in the new explanation of the second law of thermodynamics.\footnote{Klein (1973), p. 62.}
\end{quote} 
Klein stresses that Maxwell ``had already seen to the heart of the matter several years ago" when Maxwell stressed in 1871 in his \textit{Theory of Heat} that the second law had only a probabilistic validity. But this was precisely what Boltzmann had \textit{not} said in the early 1870s, and Klein himself recognizes that it was only in 1877, under pressure from Loschmidt's objection, that Boltzmann constructed a ``fully statistical interpretation of the second law''.\footnote{\textit{Ibid}, p. 63.} 

We have borrowed this terminology in the title of our paper. ``Fully statistical mechanics'' for us represents the introduction of considerations that transcend claims about the frequencies or distributions of molecules in position or velocity space for a given finite gas at a given time--- in other words, that transcend assumptions about the actual, as opposed to probable configuration or evolution of the gas. \textit{It is about the transition from counting to expecting}. Such considerations, as is well known, in turn account for the introduction into the discipline of such varied notions as infinite ensembles or subjective schemes of inductive inference. It is in going from statistical mechanics to fully statistical mechanics that the conceptual complexities and obscurities of probability truly make their appearance. 

We finish this subsection with two remarks. 

In his detailed 1978 analysis of Boltzmann's $H$-theorem, Kuhn argued that the very dynamical structure of the theorem requires a probabilistic interpretation of the density function $f$, something which Boltzmann failed to see in 1872. Kuhn drew attention to the fact that the density function is coarse-grained, and hence no deterministic equation of motion can exist for it.\footnote{Kuhn (1978), p. 45. It should be noted that Kuhn, like Emch and Liu, also argues that the SZA, in itself, has an intrinsically probabilistic meaning; see pp. 40 and 45.} He stated that ``Boltzmann's equations for the time rate of change of $f$ and $H$ determine only average or most probable values. Many other rates of change are compatible both with the given initial distribution and with mechanics.''\footnote{Kuhn (1978), p. 45.} However, Kuhn did not spell out how it is that deterministic equations are restored when the $f$ and $H$ are so interpreted in terms of probable values.\footnote{Recall the discussion at the end of section 2.1 above.} More to the point, and as we stressed earlier in the paper, a version of the $H$-theorem can be given in the wind-tree model that does not require the differentiability of $f$ or $H$ with respect to time, and it works without assigning a probabilistic interpretation to $f$.

The final point is that as Poincar\'{e} emphasized\footnote{See footnote 22 above.}, it is only by introducing a probabilistic element into arguments like the $H$-theorem, that the mechanical theory of heat have ``any chance of success'', at least in the light of the Zermelo recurrence objection. But as Poincar\'{e} hinted in 1893, and Zermelo stressed in 1896\footnote{See section 4 above.}, probability is itself not a time-asymmetric notion; in particular the Schuetz-Boltzmann notion of a fluctuating $H$-curve for the universe does not pick out a preferred direction of time.\footnote{See Uffink (2007) section 4, and Torretti (2007) p. 748 in this connection.} With respect to the two great early objections to the 1872 $H$-theorem, Boltzmann's probabilistic ploy was not sufficient to meet the objection it was originally aimed at. It seems Boltzmann himself may have realized this in 1897.\footnote{See the last remarks in section 4 above.}

\subsection{Post-$H$-theorem Boltzmann: Probability reigns}

Both in his (first) 1895 contribution to the \textit{Nature} debate and in his 1896 reply to Zermelo, Boltzmann claims that a gas approaches equilibrium because there are many more ways the gas can arrange itself in equilibrium that in any other state. This now familiar combinatorial argument, which---for better or worse---transcends any dynamical considerations, dates back to Boltzmann's 1877 work\footnote{Boltzmann (1877b).}; these early arguments have received careful critical analyses by Kuhn and Uffink.\footnote{See Kuhn  (1978) pp. 46-54, and Uffink (2007), section 4.}

Let us leave historical niceties aside, and consider the modern version of what is widely taken to be Boltzmann's new argument, wherein the SZA is essentially lost from view. Thus suppose
 we are given a typical set of measurements we can make on our dynamical system using macroscopic instruments with finite sensitivity.  This corresponds to a set $\Sigma$ of ``macrovariables'', which allow us to partition the the phase space $\Gamma$ into $\mu$-measurable subsets (``macrostates'') in the standard fashion.\footnote{Every phase point in $\Gamma$ is in exactly one macrostate, while the map $\Gamma$ to the set of macrostates is many-one.
Every macrostate corresponds to a unique set of values for the set $\Sigma$ of ``macrovariables'', and
every macrostate is invariant under permutations of identical microsystems.}
Now for every phase point $\omega \in \Gamma$, its associated macrostate $M(\omega)$ will have measure $\mu(M(\omega))$. The \textit{equilibrium} macrostate is defined to be that which has the largest measure. The \textit{Boltzmann entropy} associated with an arbitrary macrostate $M$ is defined as $S_B(M) = k\log (\mu(M))$, where $k$ is Boltzmann's constant. It will be assumed that $\mu$ is the standard Lebesgue measure on (the $\sigma$-algebra of subsets of) the phase space $\Gamma$, or rather that induced on the constant energy hypersurface $\Gamma_E$.

Boltzmann's post-$H$-theorem version of the second law of thermodynamics was essentially this:

\vspace{.2in}

\textit{Principle of probable equilibration} (PPE): Suppose that at some time $t _0$, the Boltzmann entropy $S_B(t_0)$ of the system is low compared with the entropy of the equilibrium state. Then for a later times $t > t_0$, it is highly probable that  $S_B(t) >  S_B(t_0)$.\footnote{It is assumed of course that the interval $t - t_0$ is small compared to the recurrence time of the system.}

\vspace{.2in}
For \textit{PPE} to hold, it is commonly understood that (i) the flow $T_t$ must be such that for the overwhelming majority (defined relative to the measure $\mu$) of microstates $\omega$ in the region $M_{t_0}$ corresponding to the initial low entropy macrostate, the macrostate $M_{t}(\omega)$ that results from the evolution of $\omega$ is such that $\mu (M_{t}(\omega)) \gg \mu(M_{t_0})$, and (ii) the measure $\mu$ is assigned a probabilistic significance. But claim (i) is highly non-trivial and cries out for justification; condition (ii) is clearly not enough.\footnote{See for example the discussion in Lavis (2005), section 4.2.}
It behoves us to recall Planck's 1908 remark (cited by Kuhn in 1978\footnote{Kuhn (1978), p. 54.}):
\begin{quotation}
Probability calculus can serve, if nothing is known in advance, to determine the most probable state. But it cannot serve, if an improbable [initial] state is given, to compute the following [state]. That is determined not be probability but by mechanics. To maintain that change in nature always proceeds from [states of] lower to higher probability would be totally without foundation.
\end{quotation}

\section{Acknowledgments}

We are very grateful to Giovanni Valente, Owen Maroney, and especially Jos Uffink, for spotting errors in earlier versions of this paper and for providing important clarifications and suggestions in the course of many discussions. One of us (H.R.B.) gratefully acknowledges the support of the Perimeter Institute for Theoretical Physics, where part of this work was undertaken. Research at Perimeter Institute is supported by the Government of Canada through Industry Canada and by the Province of Ontario through the Ministry of Research \& Innovation.

\section{Appendices}

\appendix
\section{Proof of the $H$-theorem analogue in the wind-tree model without differentiability}

The log sum inequality\footnote{We are grateful to Owen Maroney for bringing this inequality to our attention.} states that for non-negative numbers $a_1, a_2, \ldots, a_n$, and $b_1, b_2, \ldots, b_n$,  then
\begin{equation}
\sum_{i=1}^n a_i\log \frac{a_i}{b_i} \geq \left( \sum_{i=1}^n a_i \right) \log \frac{\sum_{i=1}^n a_i}{\sum_{i=1}^n  b_i}.
\end{equation}
with equality \textit{iff} $\frac{a_i}{b_i} =$ \textit{constant}. When $\sum_{i=1}^n a_i = \sum_{i=1}^n b_i$, we get
\begin{equation}
\sum_{i=1}^n b_i \log b_i \geq \sum_{i=1}^n b_i \log a_i.
\end{equation}
Note that these inequalities are independent of the base of the logarithm.

Now, in the interval of time $[t_0, t_0 + \Delta t]$, the $H$-function changes by 
\begin{equation}
\Delta H = \sum_{i=1}^4 \left[ f'_i \ln f'_i - f_i \ln f_i \right],
\end{equation}
where (see eqn. (5) above)
\begin{equation}
f'_i = f_i(t_0 + \Delta t) = f_i + k \Delta t( f_{i+1} + f_{i-1} - 2f_i).
\end{equation}
From (18) and (19), after rearranging we get
\begin{equation}
\Delta H = \sum_{i=1}^4 \left[ \{f_i \ln f'_i - f_i \ln f_i\} + k \Delta t ( f_{i+1} + f_{i-1} - 2f_i) \ln f'_i\right].
\end{equation}
Applying (17) twice in the second  ($k$-dependent) term in the sum, and noting that $\sum_{i=1}^4 f_{i \pm 1} \ln f_{i \pm 1} = \sum_{i=1}^4 f_i \ln f_i$, one infers that the second term is less than or equal to
\begin{equation}
2k \Delta t \sum_{i=1}^4 ( f_i \ln f_i -  f_i  \ln f'_i ).
\end{equation}
Thus
\begin{equation}
\Delta H \leq (1 - 2k \Delta t) \sum_{i=1}^4 ( f_i \ln f'_i -  f_i  \ln f_i ).
\end{equation}

Given that $k \Delta t$ is positive, using (17) again we get for sufficiently small  $k\Delta t$ the desired result
\begin{equation}
\Delta H \leq 0.
\end{equation}
Thus, if the Collision Assumption is satisfied, $H$ will be non-increasing.

We do not expect the Collision Assumption to be satisfied exactly at all times and for all intervals $\Delta t$; indeed, it \emph{cannot} be satisfied exactly unless $k \Delta t  f_i$ is integral for each $i$.  The $H$-theorem is relevant, nevertheless, because $H$ is guaranteed to be non-increasing provided that the Collision Assumption is approximately satisfied.  Define the quantities $\varepsilon_{i j}$, for $j = i \pm 1$, by
\begin{equation}
N_{ij}(\Delta t) =k \Delta t  \left( f_i +  \varepsilon_{i j} \right).
\end{equation}
Then, in place of (19), we have
\begin{equation}
f'_i = f_i + k \Delta t \left[ (f_{i +1} +f_{i -1} - 2 f_i) + (\varepsilon_{i+1, i} +  \varepsilon_{i-1, i}  - \varepsilon_{i, i+1} - \varepsilon_{i, i-1}) \right].
\end{equation}
The reasoning leading to (22) now gives us,
\begin{eqnarray}
\nonumber \Delta H &\leq& (1 - 2k \Delta t) \sum_{i=1}^4 ( f_i \ln f'_i -  f_i  \ln f_i ) 
\\
&+& k \Delta t \sum_{i=1}^4  (\varepsilon_{i+1, i} +  \varepsilon_{i-1, i} - \varepsilon_{i, i+1}  - \varepsilon_{i, i-1}) \ln f'_i.
\end{eqnarray}

Unless $f'_i = f_i$ for all $i$ (in which case $\Delta H = 0$), the first sum will be negative, and so, when  $2 k \Delta t < 1$, we will have  $\Delta H \leq 0$ if all $\varepsilon_{i j}$ are sufficiently small.  A weaker condition suffices: $\Delta H \leq 0$ if
\[
| (\varepsilon_{i+1, i} +  \varepsilon_{i-1, i}) - (\varepsilon_{i, i+1}  + \varepsilon_{i, i-1})|
\]
is sufficiently small for all $i$.

\section{More Zermelo-type proofs without recurrence}

Let $\langle \Gamma, \mathcal{A} \rangle$ be a measure space, $s: \mathcal{A} \rightarrow R$ a finitely additive function (that is, for disjoint $A, B \in \mathcal{A}, s(A \cup B) = s(A) + s(B)$), such that $s(\Gamma) < \infty$.  Let $T: \Gamma \rightarrow \Gamma$ be such that $s$ is nondecreasing under the action of $T$. That is, for all $A \in \mathcal{A}$, $s(T(A)) \geq s(A)$.

We want to show that, under mild conditions on $T$, $s$ will be constant under the action of $T$---that is,  $s(T(A)) = s(A)$ for all $A \in \mathcal{A}$. 
\bigskip

\textbf{Theorem B1.}
\textit{If $T$ is a bijection, then $s$ is constant under the action of $T$.}

\bigskip
\noindent \emph{Proof}.  Take $A \in \mathcal{A}$, and let $B = \Gamma \setminus A$.  Since $T$ is a bijection, $T(\Gamma) = \Gamma$, and $T(A) \cap T(B) = \emptyset$.  Therefore,
\begin{equation}
s(\Gamma) = s(T(A)) + s(T(B)) = s(A) + s(B).
\end{equation}
If $s(T(A))  > s(A)$, this increase must be compensated for by having  $s(T(B)) < s(B)$, contradicting the assumption that $s$ is nondecreasing.  Therefore, $s(T(A))  = s(A)$.

\bigskip

\textbf{Theorem B2.}
If there exists a measure $\mu$ on $\Gamma$ that is conserved under $T$, such that $s(A) = 0$ whenever $\mu(A) = 0$, then $s$ is constant under the action of $T$.

\bigskip
\noindent \emph{Proof}.   As before, take $A \in \mathcal{A}$, and let $B = \Gamma \setminus A$.  Since we're not assuming that $T$ is invertible, we can't conclude that  $T(A) \cap T(B) = \emptyset$, but, since $\mu$ is conserved, we \emph{can} conclude that $\mu(T(A) \cap T(B)) = 0$.  This gives us, once again,
\begin{equation}
s(\Gamma) = s(T(A)) + s(T(B)) = s(A) + s(B),
\end{equation}
and $s$ is constant under $T$.

Now if $S$ is an integrable function on $\Gamma$, for some measure $\mu$, take
\begin{equation}
s(A) = \int_A S \: d \mu.
\end{equation}
Then the theorems show that, if $S$ is almost everywhere non-decreasing, it is almost everywhere constant along the trajectories.

\section{The Zermelo proof of recurrence without continuity or invertibility}

We wish to prove the Theorem in section 5.3.1 without appeal to continuity of the flow; we assume the weaker condition that the flow is \textit{measurable}. (A function $F$ is measurable iff, for every measurable set $X$, $F^{-1}(X) = \{x \  | \ F(x) \in X\}$ is a measurable set.)

Consider as in section 5.2.2 any measurable set $g$ and its entire future development
\begin{equation}
G = \bigcup_{m \geq 0} T_{m\tau} (g).
\end{equation}
 associated with multiples of the given time period $\tau$. Similarly consider the entire future development of $T_{\tau}(g)$:
\begin{equation}
G_{\tau} = \bigcup_{m \geq 1} T_{m\tau} (g).
\end{equation}
First, we want to show that almost all points in $g$ are in $G_{\tau}$. 

\bigskip
\textbf{Lemma C1.} 
$T_{\tau}G = G_{\tau}$. 

\bigskip
\textit{Proof} Take any $x \in T_{\tau}(G)$. Then there exists a $y \in G$ such that $x = T_{\tau}y$, which means that there exists a $z \in g$, and $t \geq 0$, such that $x = T_{\tau}T_t z$. By the semi-group property of the flow, $T_{\tau}T_t = T_{\tau + t}  =T_t T_{\tau}$. So $x \in G_{\tau}$. Conversely, take any $x \in G_{\tau}$. Then there exists a $z \in g$ and $t' \geq 0$ such that $x = T_tT_{\tau} z = T_{\tau}T_t z$. Thus there exists a $y \in G$ such that $x = T_{\tau}y$. So $x \in T_{\tau}(G)$.\footnote{Here is how Zermelo (1896) put the argument.``When the region $g$ changes with time, at the same time all its `later phases' will change into the following ones, and $G$ also changes so that it represents at any instant the 'future' of the corresponding of the corresponding phase [$T_t(g)$].''}

\bigskip
Now since $\mu(\Gamma)$ is finite, $G$ has finite measure. By construction, $G_{\tau}$ is contained in $G$. But from the Lemma, and because the flow is measure-preserving, we have $\mu(G_{\tau}) = \mu(G)$. Therefore the set of elements of $G$ that are not also in $G_{\tau}$ is of measure zero. But since $g$ is contained in $G$, all but a set of measure zero of states in $g$ are in $G_{\tau}$, as desired. 

This means that there are states in $g_{\tau}$ that return to $g$, which in turn means that there are states in $g$ that return to $g$ for some time $t \geq \tau$.  Now let $r$ be the set of recurrent points in $g$. We want to show finally is that almost all states in $g$ recur, i.e. that $\mu(r) = \mu(g)$.

Let $\tilde{r}$ be the set of points in $g$ that are future destinations of points in $g$:
\begin{equation}
\tilde{r} = g \cap G_{\tau} = \bigcup_{m \geq 1} T_{m\tau}(r).
\end{equation}
From above we know that $\mu(\tilde{r}) = \mu(g)$.

Let us partition $r$ into disjoint sets of points that recur at the same time. Let $s_m$ be the set of points in $r$ whose first recurrence occurs at $m\tau$. (Note that $T_{m\tau}(s_m) \in r$ is measurable, so it follows from the assumption that $T_t$ is measurable that $s_m$ is measurable.) Then $s_m \cap s_n = \emptyset$ for distinct $m$, $n$, and 
\begin{equation}
r = \bigcup_{m\geq1} s_m.
\end{equation}
Moreover, since for every point in $\tilde{r}$ there is a time of first recurrence,
\begin{equation}
\tilde{r} = \bigcup_{m \geq  1}T_{m\tau}(s_m).
\end{equation}
Since each $s_m$ is measurable, it follows that 
\begin{equation}
\mu(r) = \sum_{m=1}^{\infty} \mu(s_m).
\end{equation}
But by the measure-preserving nature of the flow, for each $m$
\begin{equation}
\mu(T_{m\tau}(s_m)) = \mu(s_m).
\end{equation}
This gives us from (30)
\begin{equation}
\mu(\tilde{r}) \leq \sum_{m=1}^{\infty} \mu(T_{m\tau}(s_m)) =  \sum_{m=1}^{\infty}\mu(s_m) = \mu(r).
\end{equation}
Since $\mu(\tilde{r}) = \mu(g)$, this gives us $\mu(g) \leq \mu(r)$. And since $r \subseteq g$, it follows finally that $\mu(r) = \mu(g)$.

\section{Proof of the weak non-intersection lemma without invertibility}

Recall that the lemma in section 5.3.1 asserts that the sequence of time developments $T_{n\tau} (\bar{g})$ of the non-recurring set $\bar{g}$ is such that for every $n, m \geq 0$ and $n \neq m$, $T_{n\tau} (\bar{g}) \cap T_{m\tau} (\bar{g}) = \emptyset$. Here we wish to prove a weaker result without appeal to invertibility of the flow, namely that $\mu \left(T_{n\tau} (\bar{g}) \cap T_{m\tau} (\bar{g}) \right) = 0$. First we prove a second lemma.

\bigskip
\textbf{Lemma D1.} If $\mathcal{T}$ is a measure-preserving map from the phase space to itself, then for any measurable sets $X$, $Y$,
\begin{equation}
\mu (\mathcal{T}(X) \cap \mathcal{T}(Y)) = \mu(X \cap Y).
\end{equation}

\textit{Proof} We make use of the fact that 
\begin{equation}
\mathcal{T}(X) \cup \mathcal{T}(Y) = \mathcal{T}(X \cup Y).
\end{equation}
If $\mathcal{T}$ is measure-preserving, 
\begin{equation}
\mu (\mathcal{T}(X \cup Y)) = \mu(X \cup Y) = \mu(X) + \mu(Y) - \mu(X \cap Y).
\end{equation}
Also,
\begin{eqnarray}
\mu (\mathcal{T}(X) \cup \mathcal{T}(Y)) &=& \mu(\mathcal{T}(X)) + \mu(\mathcal{T}(Y)) - \mu(\mathcal{T}(X) \cap \mathcal{T}(Y))\\
&=& \mu(X) + \mu(Y) -\mu (\mathcal{T}(X) \cap \mathcal{T}(Y)).
\end{eqnarray}
Comparing XX and XX gives us
\begin{equation}
\mu (\mathcal{T}(X) \cap \mathcal{T}(Y)) = \mu (X \cap Y).
\end{equation}

\bigskip
We apply this lemma now to the Poincar\'{e} theorem by taking $\mathcal{T} = T_{n \tau}$. For all $k, n \geq 0$, 
\begin{equation}
\mu\left( T_{n \tau} (\bar{g}) \cap T_{(n+k)\tau} (\bar{g}) \right) = \mu \left( T_{n \tau} (\bar{g}) \cap T_{n \tau} \left( T_{k \tau} (\bar{g}) \right) \right) = \mu\left( \bar{g} \cap T_{k \tau} (\bar{g}) \right).
\end{equation}
Since $\bar{g} \cap T_{k \tau} (\bar{g}) = \emptyset$ for all $k > 0$, we have the result that
\begin{equation}
\mu \left( T_{n \tau} (\bar{g}) \cap T_{(n + k)\tau} (\bar{g}) \right) = 0
\end{equation}
for all $n \geq 0$, $k > 0$.

\section{The Collision Assumption is not time-symmetric}

The Collision Assumption defined in section 3 above for the wind-tree model is time-asymmetric in the sense that it is defined for pre-collision, not post-collision velocities. In this Appendix, we show furthermore that if it holds at time $t_0$ for forward velocities, and at $t_0 + \Delta t$ for the reversed velocities, then the distribution at both $t_0$ and $t_0 + \Delta t$ must be the equilibrium distribution.

We can rewrite equation (19) above in the compact form
\begin{equation}
\mathbf{f}(t_0 + \Delta t) = (I - k\Delta t C) \mathbf{f}(t_0),
\end{equation}
where $\mathbf{f} = (f_1, \ldots , f_4)$ and 
\begin{equation}
C= \left( \begin{array}{rrrr}
2 & -1 & 0 & -1 \\
-1 & 2 & -1& 0 \\
0 & -1 & 2 & -1 \\
-1 & 0 & -1 & 2 \end{array} \right).
\end{equation}

Now we can implement the time reversal operation in equation (8) at any time by the matrix equation $\mathbf{f'}(t) = P\mathbf{f}(t)$, where the permutation $P$ is given by
\begin{equation}
P = \left( \begin{array}{rrrr}
0 & 0 & 1 & 0 \\
0 & 0 & 0 & 1 \\
1 & 0 & 0 & 0 \\
0 & 1 & 0 & 0 \end{array} \right).
\end{equation}
It is easy to check that, as expected, $P^2 = I$, and furthermore that  $P$ commutes with $C$.

Now if the Collision Assumption holds for the reversed velocities at $t_0 + \Delta t$, then
\begin{equation}
P\mathbf{f}(t_0) = (I - k\Delta t C) P\mathbf{f}(t_0 + \Delta t).
\end{equation}
But acting on (46) by $P$, we get
\begin{eqnarray}
P\mathbf{f}(t_0 + \Delta t) &=& P(I - k\Delta t C) \mathbf{f}(t_0)\\
&=& (I - k\Delta t C) P\mathbf{f}(t_0).
\end{eqnarray}
And substituting $P\mathbf{f}(t_0)$ from (49) we obtain 
\begin{equation}
P\mathbf{f}(t_0 + \Delta t) = (I - k\Delta t C)^2P \mathbf{f}(t_0 + \Delta t).
\end{equation}
And this is only possible for arbitrary small $\Delta t$ if $\mathbf{f}(t_0 + \Delta t)$, and hence $\mathbf{f}(t_0)$, are the equilibrium distribution (see equation (7)) $\bar{\mathbf{f}} = \frac{1}{4}I$.

\bigskip
\noindent \textbf{\Large{References}}

\bigskip
\noindent Albert, D. Z. (2000). \textit{Time and Chance}. Cambridge, Mass: Harvard University Press.\\

\noindent Ambegaokar, V. and Clerk, A. A. (1999). Entropy and Time, \textit{American Journal of Physics, 67}, 1068-1073.\\

\noindent Bacciagaluppi, G. (2007), Probability and time symmetry in classical Markov processes, unpublished ms. 

\noindent Bricmont, J. (1995). Science of Chaos or Chaos in Science?, \textit{Physicalia, 17},  159-208. Reprinted in P. R. Gross, N. Levitt and M. W. Lewis (Eds.), \textit{The flight from science and reason}. New York: New York Academy of Sciences, pp. 131-176.\\

\noindent Boltzmann, L. (1872). Weitere Studien \"{u}ber das W\"{a}rmegleichgewicht unter Gasmolek\"{u}n, \textit{Sitzungberichte der Akademie der Wissenschaften zu Wien, mathematisch-naturwissenschaftliche Klasse, 66}, 275-370. Reprinted in Boltzmann (1909) Vol. I, pp. 316-402. English translation in Brush (2003), pp. 262-349.\\

\noindent Boltzmann, L. (1877a).   Bemerkeungen \"{u}ber einige Probleme der mechanische W\"{a}rmtheorie, \textit{Sitzungberichte der Akademie der Wissenschaften zu Wien, mathematisch-naturwissenschaftliche Klasse, 75}, 62-100. Reprinted in Boltzmann (1909) Vol. 2, pp.116-122. Partial English translation in Brush (2003), pp. 362-376 (NB: the title ascribed to this paper in Brush (2003), pp. vi, 362, and 362n., is that of Boltzmann (1877b)).\\

\noindent Boltzmann, L. (1877b).   \"{U}ber die Beziehung eines allgemeinen mechanischen Satzes zum zweiten Satze der W\"{a}rmtheorie, \textit{Sitzungberichte der Akademie der Wissenschaften zu Wien, mathematisch-naturwissenschaftliche Klasse, 76}, 373-435. Reprinted in Boltzmann (1909) Vol. 2, pp.164-223.\\

\noindent Boltzmann, L. (1895a). On certain questions of the theory of gases, \textit{Nature, 51, no. 1322}, 413-415.\\

\noindent Boltzmann, L. (1895b). On the minimum theorem in the theory of gases, \textit{Nature, 52, no. 1340}, 221.\\

\noindent Boltzmann, L. (1896). Entgegnung auf die w\"{a}rmetheoritischen Betrachtungen de Hrn. E. Zermelo, \textit{Wiedemann's Annalen 57}, 773-784; also in Boltzmann (1909), Vol. III, paper 119.\\

\noindent Boltzmann, L. (1897). Zu Hrn Zermelos Abhandlung ``\"{U}ber die mechanische Erkl\"{a}rung irreversibler Vorg\"{a}nge''. \textit{Annalen der Physik 60}, 392-398. Reprinted in Boltzmann (1909), Vol. III, pp. 579-584. English translation in Brush (2003), pp. 412-419.\\

\noindent Boltzmann, L. (1909). \textit{Wissenschaftliche Abhandlungen}, Vol. I, II, and III. F. Hasen\"{o}hrl (ed.), Leipzig: Barth. Reissued New York: Chelsea, 1969.\\

\noindent Brush, S. G. (2003). \textit{The kinetic theory of gases}, London: Imperial College Press.\\

\noindent Bryan, G. H. (1894). The Kinetic Theory of Gases, \textit{Nature, 51, no. 1312}, 176.\\

\noindent Bryan, G. H. (1913). Samuel Hawksley Burbury, 1831-1911,  
 \textit{Proceedings Royal Society of London, series A, 88, no. 606}, pp. i-iv. (The obituary notice was attributed to ``G. H. B.".)\\

\noindent Burbury, S. H. (1894). Boltzmann's Minimum Function, \textit{Nature, 51, no. 1308}, 78.\\

\noindent Burbury, S. H. (1895a). Boltzmann's Minimum Function, \textit{Nature, 51, no. 1318}, 320.\\

\noindent Burbury, S. H. (1895b). Boltzmann's Minimum Function, \textit{Nature, 52, no. 1335}, 104-105.\\

\noindent Burbury, S. H. (1899). \textit{A treatise on the kinetic theory of gases}. Cambridge: Cambridge University Press.\\

\noindent Culverwell, E. P. (1894a). Dr. Watson's proof of Boltzmann's theorem on the permanence of distributions, \textit{Nature, 50, no. 1304}, 617.\\

\noindent Culverwell, E. P. (1894b). Boltzmann's Minimum Theorem, \textit{Nature, 51, no. 1315}, 246.\\

\noindent Davies, P. C. W. (1974). \textit{The Physics of Time Asymmetry}, Belfast: Surrey University Press.\\

\noindent Dias, P. M. C. (1994). Will someone say exactly what the $H$-theorem proves? A study of Burbury's Condition A and Maxwell's Proposition II. \textit{Archive for History of Exact Sciences, 46}, 341-366.\\

\noindent Eggarter, T. P. (1973) A comment on Boltzmann's $H$-theorem and time reversal, \textit{American Journal of Physics, 41}, 874-877.\\

\noindent Ehrenfest, P. and Ehrenfest, T. (1990). \textit{The Conceptual Foundations of the
Statistical Approach in Mechanics}, Toronto: Dover Publications, Inc. This
is  a republication of the English translation published by Cornell University
Press (Ithaca, NY) in 1959. The original German publication was: Begriffliche 
Grundlagen der Statistischen Auffassung in der Mechanik, \textit{Enzyclop\"{a}die der Mathematischen Wissensschaften, 4}, F, Klein and C. M\"{u}ller (eds.); Leibzig:Teubner, pp. 3-90 (1912).\\

\noindent Emch, G. G. and Liu, C. (2002). \textit{The Logic of Thermostatistical Physics}. Heidelberg: Springer-Verlag.\\

\noindent Huang, K. (1963). \textit{Statistical Mechanics}. New York: Wiley.\\

\noindent Jeans, J. H. (1954). \textit{The dynamical theory of gases}. New York: Dover.\\

\noindent Khinchin, A. I. (1949). \textit{Mathematical Foundations of Statistical  Mechanics}. New York: Dover Publications, Inc.\\

\noindent Klein, M. J. (1973). Mechanical explanation at the end of the nineteenth century, \textit{Centaurus, 17}, 58-82.\\

\noindent Kuhn, T. S. (1978). \textit{Black-Body Theory and the Quantum Discontinuity, 1894-1912}, New York: Oxford University Press.\\

\noindent Lavis, D. A. (2005). Boltzmann and Gibbs: An attempted reconciliation, \textit{Studies in History and Philosophy of Modern Physics, 36}, 245-273.\\

\noindent Lockwood, M. (2005). \textit{The labyrinth of time: introducing the universe}. Oxford: Oxford University Press.\\

\noindent Loschmidt, J. (1876). \"{U}ber den Zustand des W\"{a}rmgleichgewichtes eines Systemes von K\"{o}rpern mit R\"{u}cksicht auf die Schwerkraft, \textit{Sitzungberichte der Akademie der Wissenschaften zu Wien, mathematisch-naturwissenschaftliche Klasse, 73}, 128-142.\\

\noindent Mackey, M. C. (2003). \textit{Time's Arrow. The Origins of Thermodynamic 
Behaviour}, New York: Dover Publications, Inc.\\

\noindent Maxwell, J. C. (1878). Tait's ``Thermodynamics'' II, \textit{Nature, Vol. XVII, no.
 431}, 278-80.\\
 
\noindent Olsen, E. T. (1993). Classical Mechanics and Entropy, \textit{Foundations of Physics Letters, 6}, 327-337.\\

\noindent Pauli, W. (1973). \textit{Pauli Lectures on Phyiscs: Volume 4. Statistical Mechanics}, C. P. Enz (Ed.). Cambridge, Mass: MIT Press.\\

\noindent Poincar\'{e}, H. (1889). Sur les tentatives d'explication mecanique des principes de la thermodynamique, \textit{Comptes rendus de l'Acad\'{e}mie des sciences} (Paris) \textit{108}, 550-553. An English translation is found in the Appendix in Olsen (1993).\\

\noindent Poincar\'{e}, H. (1890). Sur le probl\`eme des trois corps et les \'{e}quations de la dynamique. \textit{Acta Mathematica, 13}, 1-270. English translation of extracts relevant to Zermelo's recurrence argument in Brush (2003), pp. 368-376.\\

\noindent Poincar\'{e}, H. (1892a). \textit{Thermodynamique. Le\c{c}ons profess\'{e}es pendant le premier semestre 1888-89 par H. Poincar\'{e}}. Compiled by J. Blondin. Paris: G. Carr\'{e}.\\

\noindent Poincar\'{e}, H. (1892b). Poincar\'{e}'s ``Thermodynamics'', \textit{Nature, 45, no. 1169}, 485.\\

\noindent Poincar\'{e}, H. (1893). Le m\'{e}canisme et l'exp\'{e}rience, \textit{Revue de M\'{e}taphysique et de Morale, 1}, 534-537. English translation in Brush (1969), pp. 203-207.\\

\noindent Poincar\'{e}, H. (1898). On the stability of the solar system, \textit{Nature, 58, no. 1495}, 183-185. Translation of paper in \textit{Annuaire du Bureau des Longitudes}, 1898; this original paper was reprinted in Poincar\'{e} (2002), chapter V.\\

\noindent Poincar\'{e}, H. (2002). \textit{Scientific opportunism/L'Opportunisme scientifique, An Anthology}. Series: Publications des Archives Henri PoincarŽ / Publications of the Henri PoincarŽ Archives. Birkh\"{a}user.\\

\noindent Price, H. (1996). \textit{Time's arrow and Archimedes' point}. Oxford: Oxford University Press.\\

\noindent Price, H. (2006). The thermodynamic arrow: puzzles and pseudo-puzzles. In \textit{Time and Matter}, Ikaros Bigi and Martin Faessler (Eds.), pp. 209--224. Singapore: World Scientific.\\

\noindent Rechtman, R., Salcido, A. and Calles, A. (1991). The Ehrenfests' wind-tree model
and the hypothesis of molecular chaos, \textit{European Journal of Physics, 12}, 27-33 (1991). \\

\noindent Reiter, W. L. (2007). Ludwig Boltzmann: A Life of Passion, \textit{Physics in Perspective 9}, 357-374.\\

\noindent Schulman (1997) \textit{Times's arrows and quantum measurement}, Cambridge University Press, 1997.\\

\noindent Skar, L. (1993). \textit{Physics and chance: Philosophical issues in the foundations of statistical mechanics}. Cambridge: Cambridge University Press.

\noindent Tait, P. G. (1892). Poincar\'{e}'s ``Thermodynamics'', \textit{Nature, 45, no. 1167}, 439.\\
                    
\noindent ter Haar, D.  (1955). Foundations of Statistical Mechanics, \textit{Reviews of Modern Physics, 27}, 289-338.\\

\noindent Torretti, R. (2007). The problem of time's arrow historico-critically reexamined, \textit{Studies in History and Philosophy of Modern Physics, 38}, 732-756.\\

\noindent van Holtent. J. W. \& van Saarloos, W.  (1980). A generalization of the Ehrenfests' wind-tree model, \textit{European Journal of Physics} \textbf{1}(3), 149-152.\\

\noindent von Plato, J.  (1994). \textit{Creating modern probability}. Cambridge: Cambridge University Press.\\

\noindent Uffink, J.  (2004). Boltzmann's work in statistical physics, \textit{Stanford encyclopedia of philosophy} $\langle$http://plato.stanford.edu/entries/statphys-Boltzmann/$\rangle$\\

\noindent Uffink, J.  (2007). Compendium of the foundations of statistical physics. In Butterfield, J. and Earman, J. (Eds.) \textit{Handbook of the philosophy of science: Philosophy of Physics} (pp. 924-1074). Amsterdam: North Holland.\\

\noindent Watson, H. W. (1894). Boltzmann's Minimum Theorem, \textit{Nature, 51, no. 1309}, 105.\\

\noindent Zeh, H. D. (2001). \textit{The Physical Basis of the Direction of Time} (4th ed.). Berlin, Heidelberg: Springer-Verlag.\\

\noindent Zermelo, E. (1896a). Uber enien Satz der Dynamik und die mechanische W\"{a}rmetheorie, \textit{Annalen der Physik, 57}, 485-94. English translation in Brush (2003), pp. 382-391.\\

\noindent Zermelo, E. (1896b). Uber mechanische Erkl\"{a}rungen irreversibler Vorg\"{a}nge, \textit{Annalen der Physik, 59}, 4793-801. English translation in Brush (203), pp. 403-411.\\

\bibliography{reversibility.bib} 
                       
\bibliographystyle{plainnat} 

\end{document}